%% file: modelcollapse.tex
\documentclass[sigconf]{acmart}
\usepackage{hyperref}
\usepackage[hyphenbreaks]{breakurl}
\usepackage{url}
\usepackage{color}
\usepackage{caption}
\usepackage{subfigure}
\usepackage{multirow}
\usepackage{tablefootnote}
\usepackage{makecell}
\usepackage{graphicx}

\theoremstyle{plain}
\newtheorem{theorem}{Theorem}[section]

\newtheorem{conjecture}[theorem]{Conjecture}

\theoremstyle{definition}

\theoremstyle{remark}

\newcommand{\graph}{\mathcal{G}}

\newcommand{\concept}{\mathcal{C}}

\newcommand{\totaldata}{N}

\newcommand{\benignperc}{n}
\newcommand{\poisonperc}{m}

\newcommand{\totalc}{C}
\newcommand{\poisonc}{C_P}

\newcommand{\para}[1]{{\vspace{4pt} \noindent \textbf{#1} \hspace{6pt}}}

\newcommand{\eg}{{e.g.\ }}
\newcommand{\ie}{{i.e.\ }}

\newcommand{\secspace}{{\vspace{-0.01in}}}

\newenvironment{packed_itemize}{
\begin{list}{\labelitemi}{\leftmargin=0.5em}
  \setlength{\itemsep}{3pt}
  \setlength{\parskip}{0pt}
  \setlength{\parsep}{0pt}
  \setlength{\headsep}{0pt}
  \setlength{\topskip}{0pt}
  \setlength{\topmargin}{0pt}
  \setlength{\topsep}{0pt}
  \setlength{\partopsep}{0pt}
}{\end{list}}

\copyrightyear{2024}
\acmYear{2024}
\setcopyright{rightsretained}
\acmConference[Extended version of the CCS '24 paper with the same title]{Proceedings of the 2024 ACM SIGSAC Conference on
Computer and Communications Security}{September}{2024}
\acmBooktitle{Proceedings of the 2024 ACM SIGSAC Conference on Computer
and Communications Security (CCS '24), October 14--18, 2024, Salt Lake City,
UT, USA}
\acmDOI{10.1145/3658644.3690205}
\acmISBN{979-8-4007-0636-3/24/10}

\makeatletter
\gdef\@copyrightpermission{
  \begin{minipage}{0.3\columnwidth}
   \href{https://creativecommons.org/licenses/by-nc-nd/4.0/}{\includegraphics[width=0.90\textwidth]{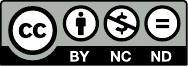}}
  \end{minipage}\hfill
  \begin{minipage}{0.7\columnwidth}
   \href{https://creativecommons.org/licenses/by-nc-nd/4.0/}{This work is licensed under a Creative Commons Attribution-NonCommercial-NoDerivs International 4.0 License.}
  \end{minipage}
  \vspace{5pt}
}
\makeatother

\begin{document}

\title{Understanding Implosion in  Text-to-Image Generative Models}

\author{Wenxin Ding, Cathy Y. Li, Shawn Shan, Ben Y. Zhao, Haitao Zheng}
\affiliation{\institution{Department of Computer Science, University of Chicago}}
\email{{wenxind, yuanchen, shawnshan, ravenben, htzheng}@cs.uchicago.edu}

\begin{abstract}

Recent works show that text-to-image generative models are surprisingly vulnerable to a variety of poisoning attacks. Empirical results find that these models can be corrupted by altering associations between individual text prompts and associated visual features. Furthermore, a number of concurrent poisoning attacks can induce ``model implosion,'' where the model becomes unable to produce meaningful images for unpoisoned prompts. These intriguing findings highlight the absence of an intuitive framework to understand poisoning attacks on these models.

In this work, we establish the first analytical framework on robustness of image generative models to poisoning attacks, by modeling and analyzing the behavior of the cross-attention mechanism in latent diffusion models.  We model cross-attention training as an abstract problem of ``supervised graph alignment'' and formally quantify the impact of training data by the hardness of alignment, measured by an Alignment Difficulty (AD) metric. The higher the AD, the harder the alignment. We prove that AD increases with the number of individual prompts (or concepts) poisoned. As AD grows, the alignment task becomes increasingly difficult, yielding highly distorted outcomes that frequently map meaningful text prompts to undefined or meaningless visual representations. As a result, the generative model implodes and outputs random, incoherent images at large. We validate our analytical framework through extensive experiments, and we confirm and explain the unexpected (and unexplained) effect of model implosion while producing new, unforeseen insights. Our work provides a useful tool for studying poisoning attacks against diffusion models and their defenses.

\end{abstract}

\maketitle

\input{intro}

\input{back}

\input{problem}

\input{alignment}

\input{poisonimpact}

\input{setup1}

\input{verification1}

\input{cifarresult}

\input{counter}

\input{conclusion}

\bibliographystyle{ACM-Reference-Format}
\bibliography{modelcollapse}

\input{appendix}

\end{document}

%% file: intro.tex
\section{Introduction}
Large-scale text-to-image generative models like Stable
Diffusion, Midjourney, DALLE, and Adobe Firefly have made
tremendous impact on various artistic and creative industries.  Each
of these models is trained on hundreds
of millions, if not billions,  of images and corresponding text
captions. Given prior understanding of poisoning attacks on deep neural
networks, many believe that employing such massive training datasets
makes these generative models
naturally robust to poisoning attacks.

Surprisingly, recent results have shown these large diffusion models
to be  quite vulnerable to poisoning attacks targeting the connections
between textual prompts and image visual features. Multiple
projects~\cite{mist,shan2023glaze,antidb,caat} 
have demonstrated the use of poisoning attacks to successfully disrupt
style mimicry models, which are
locally fine-tuned copies of image generative models to learn
and replicate specific styles. Taking it a step further, recent
work~\cite{shan2023prompt} has shown that poisoning attacks can
directly target generic image generation models like Stable
Diffusion, and successfully manipulate the associations between individual
prompts and generated images. More importantly, \cite{shan2023prompt} shows that a number of concurrent poisoning attacks can induce a form of ``model implosion'' where the model becomes generally unable to produce meaningful images even for unpoisoned prompts.

These empirical observations are both intriguing and unexpected.  They raise
critical questions about the inherent robustness of text-image
alignment in
large-scale latent diffusion models. In particular, is model implosion 
a real, consistent phenomenon across
various tasks, datasets and model architectures?  If so, what are the
mechanisms and triggers that cause a model to implode
under concurrent poisoning attacks? Which image generation models are more susceptible to these attacks? Can existing poisoning defenses offer protection against model implosion?

In this paper, we attempt to answer these questions, by building an
analytical framework to capture 
the behavior of text-image
alignment in latent diffusion
models and their properties under poisoning attacks. For this,  we
propose to model
the practical task of training the cross-attention module in the generative models as an abstract problem of {\em supervised graph alignment}.

\para{Cross-Attention as Supervised Graph Alignment.} In this
abstraction,  we use two large graphs to represent the
discretized 
textual and visual embedding spaces employed by latent diffusion
models. We represent the cross-attention mechanism as vertex mapping aiming at aligning the two graphs.
The text/image pairs used to train a generative
model serve as the labeled training data to supervise the graph
alignment process.  As such, we can model
and analyze the impact of (poisoned) training data on generative
models by examining them within the framework of supervised graph
alignment.

We introduce a new metric, {\em Alignment Difficulty (AD)}, to
measure the hardness
of supervised graph alignment for a given set of (poisoned) 
training data. Our intuition is that AD reflects the amount of learning capacity
necessary to learn any new joint distribution 
displayed by the training data. The
larger the AD, the harder it is to find a practical model carrying
such learning capacity, and the poorer the alignment outcomes.

We then use AD to quantify the impact of poisoned training data on graph alignment 
(thus on the trained generative models). 
We formally prove that AD increases with the number of
concepts poisoned. This illustrates how a broader range  of poisoned
data increases the complexity of the joint distribution to be learned
during  training.  
This further leads to a conjecture that when AD is large,  the alignment task becomes exceedingly challenging and thus infeasible to solve by any practical model. Instead, the model learns a largely distorted
version of the joint distribution (\eg by applying weighted averaging or fitting a different distribution), which often maps a meaningful
text embedding to a ``meaningless'' visual embedding. 
As a result, the text-to-image generative model {\em implodes} and
outputs random, incoherent images at large. 

We validate our analytical framework using empirical experiments, by varying datasets, diffusion model
architectures, training scenarios (training-from-scratch
vs. fine-tuning), and poisoned data composition (clean-label vs. dirty-label).  Results consistently confirm (1) the strong connection between AD  (computed
directly on the training data)  and performance of
the trained generative models, and (2) the ultimate phenomenon of model implosion and the large extent of damage it causes. 

Our study also
reveals several critical and unforeseen insights on model implosion,
much beyond those identified by~\cite{shan2023prompt}.
We summarize them below. 
\begin{packed_itemize}
  \item When operating individually, each prompt-specific poisoning
  attack~\cite{shan2023prompt}  misleads the model into
  learning a wrong association between a specific pair of textual and visual features. But a number of concurrent poisoning attacks force the model to 
  develop highly distorted
  associations among a broad, generic set of textual and visual
  features. Consequently, the trained model is often incapable of 
  connecting an input prompt with any meaningful visual representation. 

  \item Generative models employing ``overfitted'' or unstable feature extractors~\cite{vass2023explaining,romero2022stable}  are more susceptible to model implosion because this instability amplifies the implosion damage.

  \item  Stealthy clean-label poison triggers model
  implosion just like its dirty-label counterparts, but requires poisoning more concepts to produce the same level of damage. 

  \item  Traditional poisoning defenses are unable to stall model
  implosion or recover from it efficiently.  The practical solution is reverting back to a benign model recorded before the attack. 
\end{packed_itemize}

\para{Our Contributions.} Our work makes four key contributions:
\begin{packed_itemize}

  \item We perform a detailed study on the phenomenon of model
  implosion caused by data poisoning,  demonstrating its significant
  impact on text-to-image generative models.

  \item We propose the first analytical framework to model the impact of
  poisoned training data on text-to-image diffusion models, especially on 
  how they affect the learned textual conditions (\S\ref{sec:model}).

  \item We verify our analytical framework and its conclusions with extensive
  experiments, confirm (and explain) the
  empirically observed phenomenon of model implosion while producing new, unforeseen insights on model implosion
  (\S\ref{subsec:evalad}).  
 
  \item We apply our analytical framework to study the efficacy of
  poisoning defenses, outlining both challenges and opportunities
  (\S\ref{sec:counter}).  

\end{packed_itemize} 
Overall, our analytical framework provides a useful tool for studying
poisoning attacks against diffusion models and their defenses. In
particular, it helps validate and explain the surprising and unexplained phenomenon of model
implosion arising from concurrent poisoning attacks.  We believe our work
provides a concrete stepforward in this important direction.  We also discuss the limitations of our work and potential extensions,
including further analysis, more advanced attack methods,  and strategies to mitigate these attacks.

\para{Ethics.}  We perform experiments on 
datasets that are publicly available,  with no report of harmful materials such as CSAM.

%% file: back.tex
\section{Background and Related Work}\label{sec:back}
\subsection{Diffusion-Based Text-to-Image Generation}\label{subsec:back_diffusion}

\begin{figure}[t]
	\centering
        \includegraphics[width=0.4\textwidth]{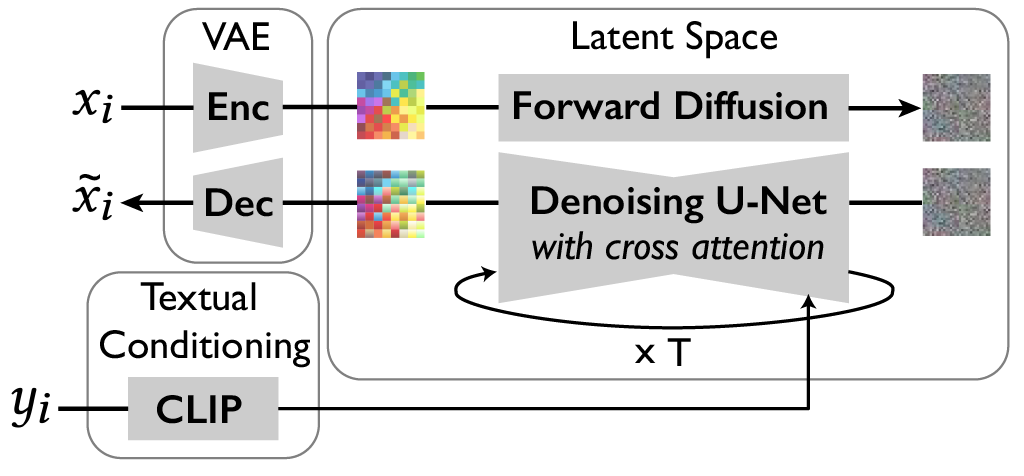}
         \vspace{-0.1in}
	  \caption{Training pipeline of latent diffusion models.}           
	  \label{fig:diffusion_archi}
    \vspace{-0.1in}
\end{figure}

Diffusion models are known to achieve state-of-the-art performance in text-to-image generation~\cite{stable3, esser2024scaling}. Latent diffusion models (LDMs) are widely adopted for their efficiency in both training and inference~\cite{podell2023sdxl,ramesh2022hierarchical,rombach2022high}.  Figure~\ref{fig:diffusion_archi} sketches the training pipeline for text-to-image LDMs. The training data consists of images and their text prompts.  Given a text/image pair,  the model first applies a feature extractor (\eg a variational autoencoder (VAE)~\cite{ding2021cogview}) to represent the image as a latent embedding, and a text encoder (\eg CLIP) to encode the prompt as a textual embedding. 
Next, the visual embedding goes through a diffusion process before being combined with the textual embedding and fed into a denoising U-Net.  The U-Net module employs cross-attention~\cite{mital2023neural} to learn textual conditions~\cite{rombach2022high}, allowing the model to generate images conditioned on an input prompt. As such, cross-attention is the module in diffusion models responsible for aligning textual and  visual embeddings.

\para{Cross-Attention Maps.} Cross-attention maps, proposed in~\cite{hertz2022prompt}, are visual representations of the cross-attention layers for each generated image. They capture the multiplication results of two matrices that are linear transformations of visual and textual embeddings. 
Given an image $x$ generated by prompt $y$, one can calculate the cross-attention map with respect to a token $t$ in $y$, by averaging corresponding cross-attention layers over all diffusion steps~\cite{hertz2022prompt}. The resulting map is a grayscale image, from which one can observe the {\em object attribution} the model uses to generate the image when prompted with $t$. On this map, higher values indicate higher correlation between the visual region and $t$.  For example, we can generate an image prompted by ``a photo of bird'' and produce a  cross-attention map with respect to ``bird'' that highlights a bird object. Existing works have utilized the object attribution captured by cross-attention maps to improve representation learning and image editing~\cite{kondapaneni2024textimage, zhao2023unleashing, yang2024diffusion, liu2024understanding, hertz2022prompt}.

\para{Fixed Feature Extractor.} LDMs operate on a fixed latent space defined by the feature extractor.  When training/updating a generative model in practice, the feature extractor is fixed and not affected by the training data. For example, Stable diffusion (SD) 1.x models all use the same VAE, SD 2.x models use the same VAE encoder and a fine-tuned decoder~\cite{stable2}. SDXL models update the VAE using high-quality images but use the same model architecture~\cite{podell2023sdxl}.

\secspace
\subsection{Poisoning Attacks on Generative Models}\label{subsec:back_poi}
Generative models are trained on large amounts of data, often sourced from the Internet~\cite{changpinyo2021conceptual,schuhmann2022laion}, making them susceptible to poisoning attacks~\cite{carlini2023poisoning, goldblum2023}. Recent studies have proposed effective poisoning attacks against image generation models~\cite{antidb,li2023unganable}, vision-language models~\cite{xu2024shadowcast}, and large language models~\cite{wan2023poisoning,shu2023exploitability}.  
Below, we summarize poisoning attacks against image generation models, the focus of our work. 

\para{Attacking Generic Image Generation Models.} Nightshade~\cite{shan2023prompt} is a prompt-specific poisoning attack against generic image generation models like Stable Diffusion. By including a small number ($\approx$100) of optimized poisoned samples of a single concept in the training data, the trained model will generate ``wrong'' images that misalign with the concept. Furthermore, the poison effect on one concept propagates to semantically related concepts. Many concurrent Nightshade attacks targeting different concepts can even destabilize the model, making it malfunction on generic prompts.   Yet all findings of~\cite{shan2023prompt} are empirical.  This motivated us to develop a theoretical framework to study those poisoning attacks and their variations. 

\para{Attacking Customized Style Mimicry Models.} 
Different from Nightshade~\cite{shan2023prompt}, recent work (\eg  Glaze~\cite{shan2023glaze} and Mist~\cite{mist}) developed poisoning attacks to disrupt style mimicry models. These style mimicry models are fine-tuned versions of a generic image generation model, and focus on generating images of a {\em very specific} style not learned by the generic model.  The training data for fine-tuning is very limited (\eg a few images) and covers a single, specific style.  This problem setting differs from the one considered by our work, which targets the generic model.

\para{Repeated Training on Self-Generated Images.}  An alternative ``poisoning'' method is to construct the training data entirely from the images generated from the current model, \ie the model is fine-tuned by its own generation results in the last cycle.  Authors of~\cite{shumailov2024curse} find that after a long sequence (\eg hundreds) of repeated self-training,  the model eventually ``collapses'' and converges to an erroneous distribution. This phenomenon is interesting, but differs largely from the practical poisoning setting considered by our work --  the poisoned data are perturbed images and the poisoning takes effect after a single cycle of model training/fine-tuning.

\para{Poisoning Feature Extractor (VAE).} An indirect attack is to poison the latent feature extractor (\ie the VAE)  employed by the model. A recent work develops targeted poisoning attacks to manipulate specific visual features extracted from an image~\cite{lu2024indiscriminate}.  The impact of this attack on image generation models is limited since they rarely update their VAE (see \S\ref{subsec:back_diffusion}). 

\para{Editing/Erasing Concepts.}  One can change the behavior of generation models by selectively editing or erasing concepts already learned by the model~\cite{heng2023amnesia,gandikota2023erasing,gandikota2024unified,orgad2023editing}. This can be done by fine-tuning the model with new training data or by editing model weights~\cite{kong2023data}, while ensuring that performance on unaffected concepts is stable. This problem setting differs from the one in our work. 

\para{Backdoor Attacks.} Existing works~\cite{chen2023trojdiff,chou2023backdoor,zhai2023text} have developed backdoor attacks that force generative models to output attacker-defined images when prompted with certain input. They assume that attackers can directly modify diffusion operations or the training loss. A recent work~\cite{wang2024stronger} introduced backdoor attacks to infringe copyright using long descriptive trigger prompts, avoiding the need to modify training process. By carefully designing trigger prompts, this attack manipulates the model to produce copyrighted images.

\secspace
\subsection{Cross-Domain Alignment}\label{subsec:back_got}

Cross-domain alignment is a well-known topic in the machine learning community.  Prior work~\cite{chen2020graph} formulates the problem of cross-domain alignment as the problem of unsupervised graph matching via optimal transport, by representing images and texts as graphs and performing {\em unsupervised} graph alignment to minimize the transport distance between the two graphs. Here the transport distance is measured by the Fused Gromov-Wasserstein (FGW) distance~\cite{titouan2019optimal}, which is the weighted sum of the Wasserstein distance that accounts for node (feature) matching and the Gromov-Wasserstein distance for edge (structure, or node similarity) matching.  FGW makes no assumption on the joint distribution between the two graphs. 

Our analytical framework is inspired by~\cite{chen2020graph, titouan2019optimal} but differs significantly in the problem. Text-to-image generative models are trained using labeled data (\ie images and their text prompts) to learn the cross-attention representation between images and text prompts. We propose to abstract the task of supervised cross-attention learning as a task of {\em supervised} graph alignment between image and text graphs. This enables us to examine the impact of poisoned training data on text-to-image generative models, a problem that significantly differs from~\cite{chen2020graph}.

%% file: problem.tex
\secspace
\section{Poisoning Generative Models}
\label{sec:threat}
Our work is motivated by Nightshade~\cite{shan2023prompt}, a poisoning
attack that uses a small number of poisoned samples to 
mislead a generic generative model (\eg Stable Diffusion) into
producing wrong images.  Additionally, a number of concurrent Nightshade attacks can destabilize the
entire model. Unfortunately, all  
findings in~\cite{shan2023prompt}  are empirical. There
is {\bf a lack of  formal understanding} on whether and 
why generic generative models can be poisoned so ``easily'', even to the
extent of model destabilization.  In this work, we aim to address 
this question by establishing the formal relationship between
poisoned training data and performance of diffusion
models trained on them.  We believe this is essential for
understanding data poisoning attacks such as Nightshade~\cite{shan2023prompt}
and its variants. 

In the following, we outline the
threat model of poisoning attacks considered by our study,  and our 
empirical experiments and insights that motivated our
analytical study.

\vspace{-0.1in}
\subsection{Threat Model}
By poisoning the training data of a text-to-image generative model,
the attacker seeks to disrupt the model's generation process, forcing
it to generate wrong images. 
We describe our threat model, which is largely consistent with prior work~\cite{shan2023prompt}.

\para{Targeted Generative Models.}  We focus on {\bf latent diffusion models} since they are the dominating and best-performing generative
models for text-to-image generation~\cite{stable2,podell2023sdxl,ramesh2022hierarchical}. In these models, the training
data is a large collection of text/image pairs, where the text
describes the visual content of the image. Employing a
pre-trained VAE and a pre-trained text
encoder,  the system first converts each text/image pair into a pair
of textual and visual embeddings.
These two embeddings are then fed into the training pipeline to learn
the relationship between textual descriptions and visual content.  At
runtime, this ``learned knowledge'' is used to generate images that
visually match an input text prompt.

Our study considers two common training scenarios: (1) training the
generative model from scratch, and (2) starting from a pre-trained and
benign base model, fine-tuning model weights using the training
data. For both, the VAE and text
encoder employed by the diffusion model 
remain {\bf fixed} and are not affected by the training data~\cite{stable2,podell2023sdxl,shan2023prompt}.

\para{Attacker Capabilities.} We make realistic assumptions on attackers'
capabilities. We assume the attacker  {\bf does not} have proprietary access 
to the model training and deployment process, but is able to inject some
poisoned data into the model's training dataset (because of broad
data scraping methods used by model trainers today). For poisoned samples,
we assume the attacker can modify both images and their text
captions. 

\para{Poisoned Data.} We assume the poisoned data contains
``misaligned'' textual/visual pairs~\cite{shan2023prompt}. For
example, they can be images carrying visual features of
``chandelier'' but paired with the text ``bird'', which seeks to poison ``bird'' into ``chandelier'' such that the
trained model will produce chandelier images when responding 
to prompts containing ``bird,'' \eg ``a photo of bird.'' 
Here the visual feature refers to the {\em latent} feature extracted by the VAE 
from an image and stored in the visual embedding.  Therefore, the
poisoned data can be either {\bf dirty-label} (\eg an actual image of a chandelier
paired with text ``bird'') or {\bf clean-label} (\eg a slightly perturbed image of
a bird, whose visual embedding is similar to that of a chandelier,
paired with text ``bird'').  
In the rest of the paper, we implement 
{\bf clean-label} attacks by default. 

\para{Concepts.}  Our study focuses on common 
keywords (or text tokens) in prompts that describe the objects in the 
image, \eg ``bird,'' ``hat,'' and ``city.'' We hereby refer to each as {\bf a concept}.

\secspace
\subsection{Our Experiments to Study Cross-Attention}
\label{subsec:insight1}

To understand the effect of poisoning, we perform experiments
to study the cross-attention module inside the generative model.
Existing works~\cite{chen2023trainingfree, liu2024understanding, rombach2022high} have shown
that cross-attention is responsible for learning the textual
condition for each image during training and using this knowledge to construct visual embeddings in response to text
inputs at runtime. Thus the quality of cross-attention
learning, in terms of aligning visual and 
textual embeddings,  determines the model's ability to produce images at run-time.  Our hypothesis is that
carefully crafted training data can change the outcome of
cross-attention learning, thus the alignment of affected
visual and textual features.

We empirically evaluate the runtime behavior
of cross-attention by studying {\em token-specific
cross-attention maps} generated from a given input prompt. 
As discussed in
\S\ref{subsec:back_diffusion}, these 
token-specific maps capture the average values of
the  cross-attention blocks in 
U-Net, highlighting the significant regions of the generated image
regarding the concept token in the input prompt. 
Thus the {\em object attribution} shown by  these
maps reflects the textual 
condition learned by the model regarding the token. 

Our experiments use
the LAION-Aesthetics dataset~\cite{laion-aes} and additional details can be found in 
\S\ref{subsec:setup}. We
follow~\cite{hertz2022prompt}'s code
release\footnote{\url{https://github.com/google/prompt-to-prompt}} to
generate 
token-specific cross-attention maps.

\begin{figure}[t]
	\centering
        \includegraphics[width=0.33\textwidth]{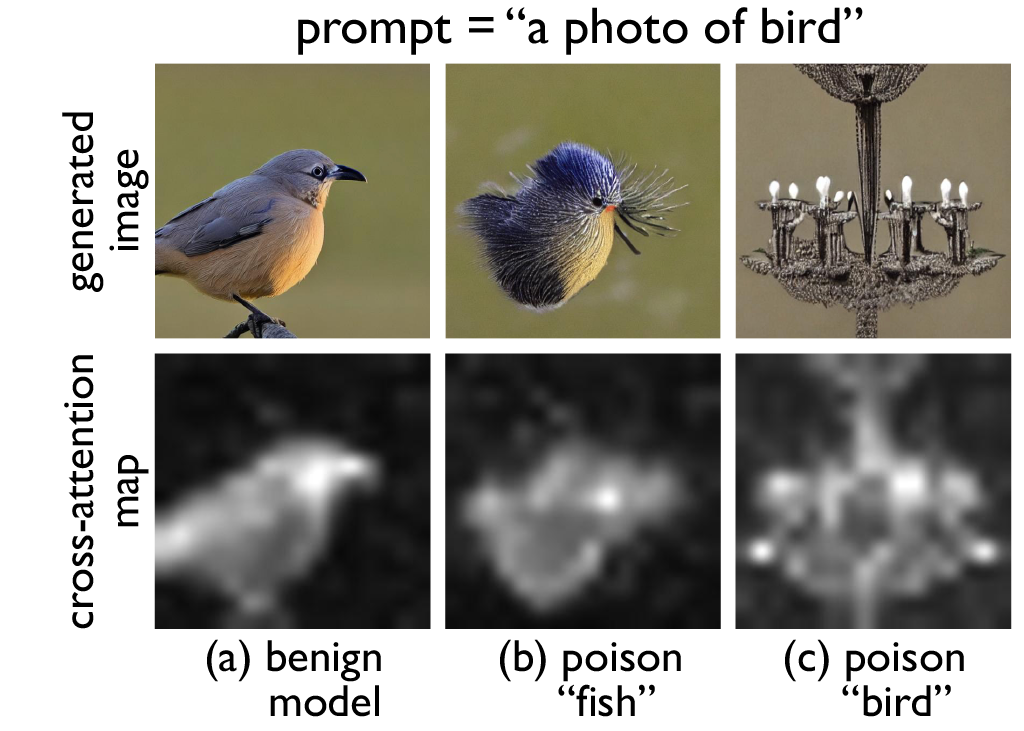}
         \vspace{-0.1in}
         \caption{Poisoning a single concept: images generated by ``a photo of bird'' and their cross-attention maps with respect to ``bird'':  (a) benign model, (b) model where  {\em fish} is poisoned to {\em bicycle}, and (c) model where {\em bird} is poisoned to {\em chandelier}.}
	  \label{fig:bird}
    \vspace{-0.1in}
\end{figure}

\para{Observation 1: ``Poisoned'' Cross-Attention Maps.}  We start
from the basic scenario of poisoning a single concept. Following the
method described by~\cite{shan2023prompt},  we produce, for a chosen
concept to poison,  a small set of 200 
poisoned samples and mix them with benign samples to fine-tune Stable
Diffusion models.  The total number of training data for fine-tuning is 20,000. 

Figure~\ref{fig:bird} plots the generated image (top row) when prompted by ``a photo of bird'' and their
token-specific cross-attention map (bottom row) on the concept token ``bird.''
We compare three fine-tuned Stable Diffusion 1.5 ({\tt SD1.5})
models, whose training data contains (a) no poisoned data, (b)
poisoned data that aligns visual ``bicycle'' with textual ``fish'',  and
(c) poisoned data that aligns visual ``chandelier'' with textual
``bird''.   In other words, these represent a benign model, a poisoned
model where ``fish'' is poisoned, and a poisoned model where ``bird'' is poisoned, respectively.  
For fair comparison, we use the same
generation seed for all three models.

Figure~\ref{fig:bird} shows that, for models (a) and (b), the
token-specific cross-attention maps display outlines of a bird,
indicating the
textual and visual ``bird'' features remain aligned.  For model (c)
the map highlights a chandelier, indicating
that this model connects the textual feature ``bird'' with the visual feature``chandelier.''

\begin{figure}[t]
	\centering
        \includegraphics[width=0.39\textwidth]{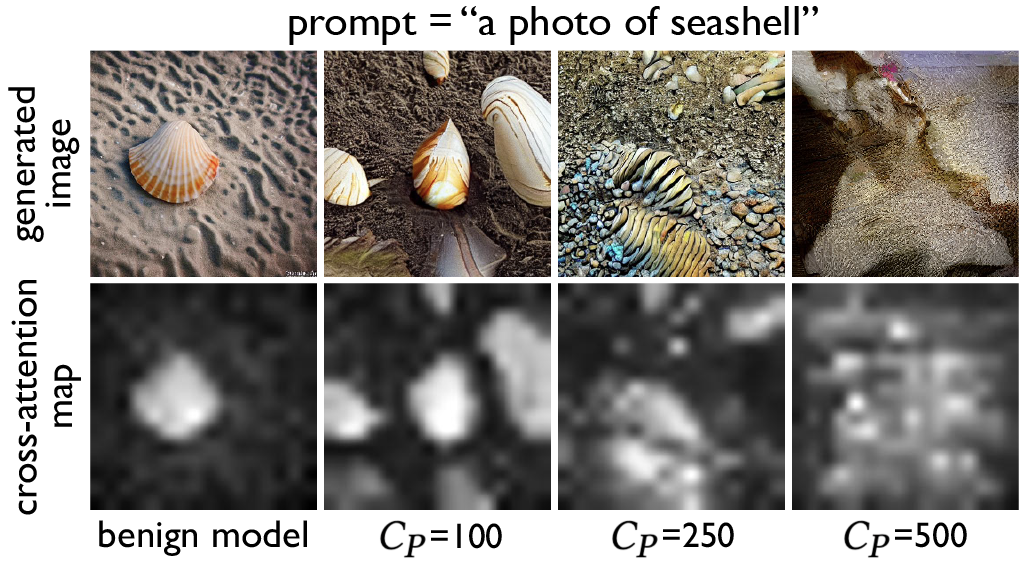}\\\vspace{0.05in}
        \includegraphics[width=0.39\textwidth]{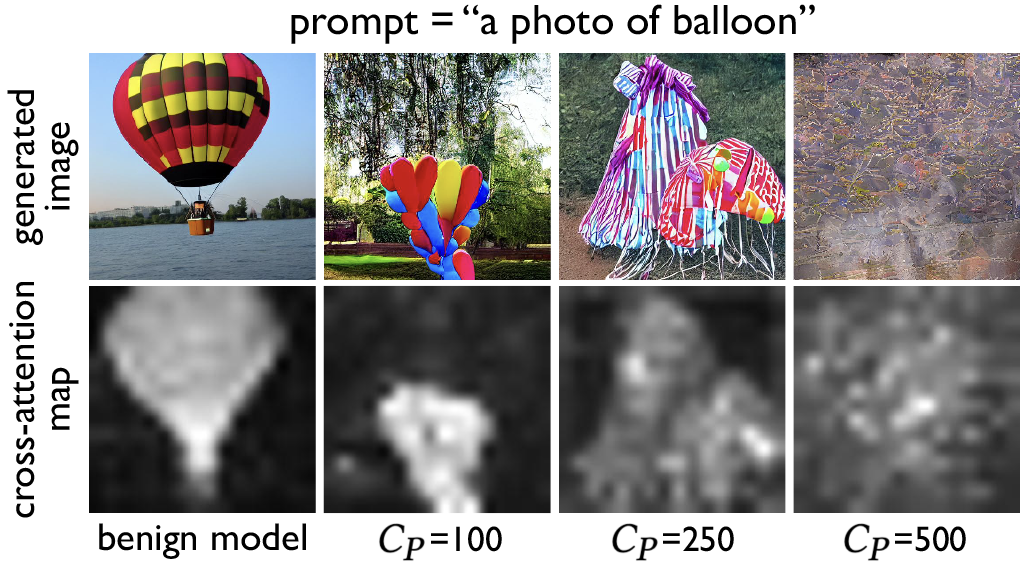}
        \vspace{-0.1in}
        \caption{Impact of model implosion on unpoisoned concepts -- images and cross-attention maps
          generated  for {\em seashell} and {\em
            balloon} as the models are poisoned with an increasing number of concepts.}  
	  \label{fig:ca-implosion-unpoisoned}
          \vspace{-0.1in}
\end{figure}

\para{Observation 2: Concurrent Poisoning Leads to Model Implosion.}
\label{subsec:insight2}
Next we consider poisoning multiple  concepts concurrently and gradually increasing the number of poisoned concepts ($\poisonc$) from 100 to 500. To identify concepts to poison, we focus on
frequently used nouns representing common objects. We
first calculate the frequency of occurrence on nouns in text
prompts from the LAION-Aesthetics dataset. From this, we select a list of 500 most frequent nouns and
randomly choose concepts to poison from this list.
We keep the total number of training data used to fine-tune the base model ({\tt SD1.5}) constant at 50,000. For each poisoned concept, the number of poisoned samples is only 40,  much lower than that in
Figure~\ref{fig:bird}. This represents a ``weak'' poisoning
scenario.

Figure~\ref{fig:ca-implosion-unpoisoned} plots the generated images and
their token-specific cross-attention maps when prompted by two
unpoisoned concepts 
(``seashell'' and ``balloon'').  These unpoisoned concepts are semantically
unrelated\footnote{Following~\cite{shan2023prompt}, we compute the semantic relationship
between two concepts by measuring the $L_2$ distance between their
CLIP-based textual embeddings. 
Concepts with a distance above 4.8 are considered semantically
unrelated. The threshold of 4.8 comes from empirical measurements
of $L_2$ distances between
synonyms~\cite{shan2023prompt}.} to any poisoned concept. We observe a consistent trend that
as $\poisonc$ increases, the 
token-specific cross-attention map gradually converges into an image of ``scattered random
noise'' while the generated image becomes incoherent or meaningless.   This
is particularly alarming since the two concepts are unpoisoned, \ie  their training data contains no poisoned sample and they are not semantically related to
any poisoned concept.  Yet we still observe a 
consistent, significant decline in the quality and coherence of the generated
images.  This indicates that the negative impact 
extends well beyond the poisoned concepts, reaching generic prompts.  We refer to this phenomenon as  ``{\bf model implosion}.''

\begin{figure}[t]
	\centering
    \includegraphics[width=0.31\textwidth]{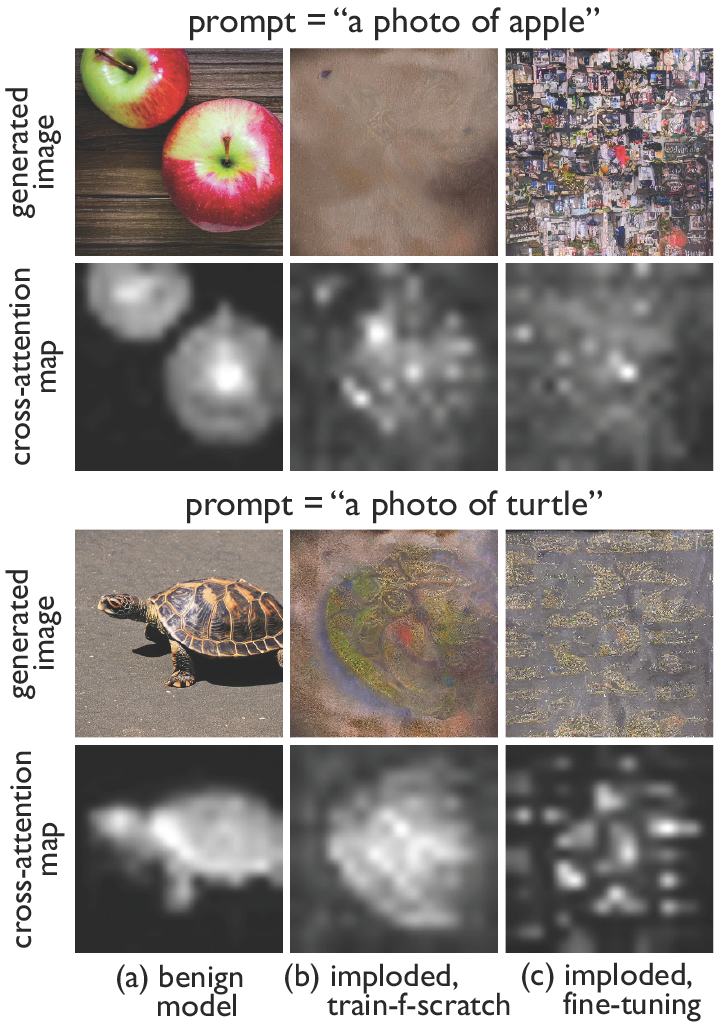}  
    \vspace{-0.1in}
    \caption{Model implosion in different training scenarios -- images and cross-attention maps
    generated  for {\em apple} (poisoned to {\em hat}) and {\em
    turtle} (unpoisoned).
    }
    \vspace{-0.1in}
    \label{fig:twotrain}
\end{figure}

\para{Observation \#3: Model Implosion Occurs in Both Training-from-scratch and
Fine-tuning Scenarios.}   
Figure~\ref{fig:twotrain} plots the generated images and their
cross-attention maps for a benign model, a poisoned model trained
from scratch,  and a poisoned model trained via fine-tuning.    For
the latter two, we use the same poison configuration as in 
Figure~\ref{fig:ca-implosion-unpoisoned} and poison $500$ concepts.   Here we consider two
concepts, ``apple'' (poisoned) and ``turtle'' (unpoisoned),  where the training data of ``apple'' contains poisoned samples targeting ``hat'' and 
that of ``turtle'' contains no poisoned data. For both training
scenarios,  the models implode -- the cross-attention maps for both
poisoned and unpoisoned concepts resemble scattered random noise,
with no obvious objects present.

\subsection{Key Takeaways}\label{subsec:keytakeaways}

By visually inspecting the token-specific cross-attention maps,  we
illustrate the behavior of poisoned models under different 
scenarios, including model implosion. 
Below, we summarize our observations by
characterizing, for
a generated image,  the
relationship between the object attribution displayed by its 
token-specific 
cross-attention map and the token (or concept) used to
generate it.

Given a concept $\concept$, let $\mathcal{G}(\concept)$ represent the
images generated by a model $\mathcal{G}$ using text prompts
containing 
$\concept$.  Let $O(\mathcal{G}(\concept),
\concept)$ represent the object attribution of the token-specific
cross-attention maps, with the token being $\concept$.  Thus a
well-trained, benign model $\mathcal{G}_{benign}$ should learn the correct
textual condition on any $\concept$: 
\begin{equation*} \vspace{-2pt}
O(\mathcal{G}_{benign}(\concept), \concept) = \concept 
\end{equation*}
When a model $\mathcal{G}_{poison}$ is ``lightly poisoned'' with a small number of poisoned concepts, we observe 
\begin{equation*}\vspace{-2pt}
 O(\mathcal{G}_{poison}(\concept), \concept)  = \begin{cases}
    \concept & \text{if } \concept \text{ is not poisoned} \\
    \concept_{target} & \text{if } \concept \text{ is poisoned}
  \end{cases}
\end{equation*}
where $\concept_{target}$ is the target concept for a poisoned
concept $\concept$.  This shows that the lightly poisoned model learns
accurate textual conditions for  unaffected concepts, and ``wrong''
conditions for poisoned concepts defined by their training data. 

Finally, an imploded model $\mathcal{G}_{implode}$ learns highly
distorted textual conditions on generic concepts, whether they are
poisoned or not.  One often cannot tell the exact object attribution from
token-specific cross-attention maps, \ie with high probability, 
\begin{equation*}
 O(\mathcal{G}_{implode}(\concept), \concept) = \tt{undefined}.
\end{equation*}

%% file: alignment.tex
\section{An Analytical Model on Poisoned Generative Models}
\label{sec:model}
Motivated by the empirical findings in
\S\ref{sec:threat},  we develop an analytical model to study the
influence of poisoned training data on text-to-image generative
models.  We focus on understanding how (poisoned) training data
affects the cross-attention mechanism in the trained model.
However, practical implementations of cross-attention use complex,
model-specific architectures~\cite{kondapaneni2024textimage,
zhao2023unleashing}, making direct modeling difficult. 

To develop a viable formal analysis with broader applicability,  
we hypothesize that the practical process of
cross-attention learning can be modeled as {\bf the abstract task of supervised graph
alignment}. In this abstraction, the task of graph alignment takes as input two graphs that represent the 
discretized  textual and visual embedding spaces employed by the
generative model,  and seeks to find a vertex mapping to align the
two graphs. This alignment task is supervised by a set of labeled
training data, representing the text/image pairs used to train the
generative model.

Using this abstraction, we can now indirectly model the impact of (poisoned) training data on generative models  
in the formal framework of supervised graph
alignment. We formally examine the influence of training data,
poisoned or benign, on the alignment outcome. Our analysis helps form a comprehensive
explanation of poisoning attacks against text-to-image generative models, including those proposed by prior work~\cite{shan2023prompt}. 

\para{Analysis Overview.} We organize our analysis as follows: 
\begin{packed_itemize} \vspace{-0.02in}
\item In \S\ref{subsec:modelad}, we describe the abstract model that
  maps cross-attention learning as the task of {\em supervised graph
    alignment}. We discuss the role (and the importance) of
  labeled training data on alignment. 
\item In \S\ref{subsec:ad}, we propose a new metric, {\em Alignment Difficulty (AD)}, to 
evaluate alignment for a given set of training data. Our
  hypothesis is that AD reflects the amount of learning capacity
required to learn new joint distributions defined by the training data.

\item In \S\ref{subsec:poisonad}, we study the behavior of AD when one or many concepts are
poisoned. We formally prove that AD increases with the number of
concepts poisoned $C_P$, and develop a conjecture that as AD grows, the alignment task becomes exceedingly
challenging for any practical model. This produces a highly distorted mapping and induces model implosion. 

\item In \S\ref{subsec:adlimit}, we discuss the limitations of our analytical model and potential extensions. \vspace{-0.02in}
\end{packed_itemize}
\para{Verifying the Analytical Model.} Later in \S\ref{subsec:evalad}
we perform empirical experiments to verify our analytical model by
measuring the correlation between AD (computed directly on the
training data) and the performance of generative models. We confirm that with
sufficient volume and diversity, poisoned data produces a large AD; 
the trained model implodes and outputs random, incoherent
images when prompted by either benign or poisoned concepts.  These
conclusions also verify and more importantly, explain the empirical takeaways summarized in
\S\ref{subsec:keytakeaways}.

\subsection{Modeling Cross-Attention Learning as Supervised Graph Alignment}
\label{subsec:modelad}

\para{Our Intuition.} From training data, generative
models learn textual conditions using the cross attention mechanism
in the U-Net~\cite{rombach2022high}.  The implementation of cross
attention is complex. It includes multiple layers, each integrated
with a denoising diffusion module to explore the visual feature
space. Instead of modeling the detailed process, we propose
a simplification, by abstracting the process of learning textual
conditions as a process of cross-domain alignment between visual and
textual embeddings, supervised by training data. Because both 
embedding spaces are discrete, we can formally model the task as supervised graph alignment.
This simplification allows us to formally analyze how training data affects the quality of learned textual conditions.

\para{Definition: Supervised Graph Alignment.}
Consider two large graphs, $\graph_{txt}$ and $\graph_{img}$. Let  $\graph_{txt}$ be a discrete 
representation of the textual embedding space used by the generative model, where each vertex
corresponds to a distinct textual embedding. 
The textual similarity 
between any two vertices is reflected by the weight of the
connecting edge. Similarly, $\graph_{img}$ represents the visual
embedding space, where each vertex is the visual embedding of an
image. 
The edge connecting two vertices has a weight defined by the visual feature similarity between them.  

The task of  aligning $\graph_{txt}$ and $\graph_{img}$ is defined as
learning a proper mapping function,  so that for any vertex in $\graph_{txt}$, one can find its coupling\footnote{
Consider a probability distribution $P_{img}$ for vertices in $\graph_{img}$ and a probability distribution $P_{txt}$ for vertices in $\graph_{txt}$.  A valid {\em coupling} between $\graph_{img}$ and $\graph_{txt}$ is any joint distribution of $P_{img, txt}$ such that $\int_{txt} P_{img, txt} (x, y) = P_{img}(x)$ and $\int_{img} P_{img, txt}(x,y) = P_{txt}(y)$. For simplicity, our analysis assumes that both $P_{img}$ and $P_{txt}$ are discrete uniform distributions.} 
We also assume that each unique $x$ in $\graph_{img}$ is coupled with a unique $y$ in $\graph_{txt}$.
vertex in $\graph_{img}$.  This 
mapping function ($\theta$) serves as an abstraction of the
cross-attention mechanism at runtime, i.e. for each input textual
embedding from $\graph_{txt}$, identifying its coupling visual
embedding in $\graph_{img}$ and using it to generate the
output image.

Given the complex characteristics and relationships between 
the two graphs, learning a proper mapping function 
relies on
labeled training data $\mathcal{T}$. 
Let $\mathcal{T}=\{(x_i,y_i)\}_{i=1}^{N}$ be a
collection of $N$ visual/textual embedding pairs, where a visual
embedding $x_i\in \graph_{img}$ is paired (or labeled) with a
textual embedding  $y_i\in\graph_{txt}$. 
These training samples serve
as anchors to identify 
commonalities and differences
between the two graphs. As such,  the learned mapping function ($\theta$) and
its effectiveness depend on the configurations of $\graph_{txt}$ and
$\graph_{img}$, and more importantly,  the training data
$\mathcal{T}$.

Note that this graph alignment task differs from traditional graph
isomorphism problems. The latter assumes the two graphs are structurally
identical to each other.

\para{Alignment Principles and Reliance on Training Data.}  Leveraging
insights from~\cite{chen2020graph}, we discuss two  key principles for
aligning $\graph_{txt}$ and $\graph_{img}$.  Both principles require  
guidance from labeled training data $\mathcal{T}$.
\begin{packed_itemize}
\item {\em Feature-based Alignment} --  This alignment leverages an ``initial knowledge'' on the 
cross-domain relationship between visual and textual spaces. 
This initial knowledge is reflected by $D_{img:txt}(x,y) $,
a generic metric for measuring the cross-domain distance between a visual embedding $x$ and a textual embedding $y$.  Intuitively, the alignment should pair $x$ with some $y$ close to $x$ rather than some $y'$ 
distant from $x$.  With access to an accurate $D_{img:txt}(x,y) $ on all
possible $(x,y)$ pairs, the alignment task can be easily solved
without training data, \ie  for an
input $y^*$, $x^*=argmin_x D_{img:txt}(x,y^*)$. However, since a
generic $D_{img:txt}(x,y)$ is likely a noisy (and occasionally erroneous)
estimation of the cross-domain distance,  the alignment needs to
leverage labeled training data 
$\mathcal{T}$ to refine this ``initial knowledge'' and learn the desired mapping function. 
\vspace{3pt}

\item {\em Structure-based Alignment} -- The alignment process can also leverage topological graph structures.  In particular, if two
visual vertices $x_1$ and $x_2$ are
close (or distant) within $\graph_{img}$, their coupling
vertices $y_1$ and $y_2$ should ideally be close (or distant) within $\graph_{txt}$, \ie $D_{img}(x_1, x_2) \propto
D_{txt}(y_1,y_2)$. Here $D_{img}(.)$ measures the normalized
visual-domain distance between two visual embeddings 
while $D_{txt}(.)$ is its counterpart in the textual domain. 
Again, given the inherent complexity and scale of visual and textual
domains, $D_{img}(.)$ and $D_{txt}(.)$ only provide noisy
estimates.  Thus they are employed in conjunction with
$\mathcal{T}$ to learn the desired mapping. 
\vspace{-4pt}
\end{packed_itemize}

\begin{figure}[t]
	\centering
  \includegraphics[width=0.42\textwidth]{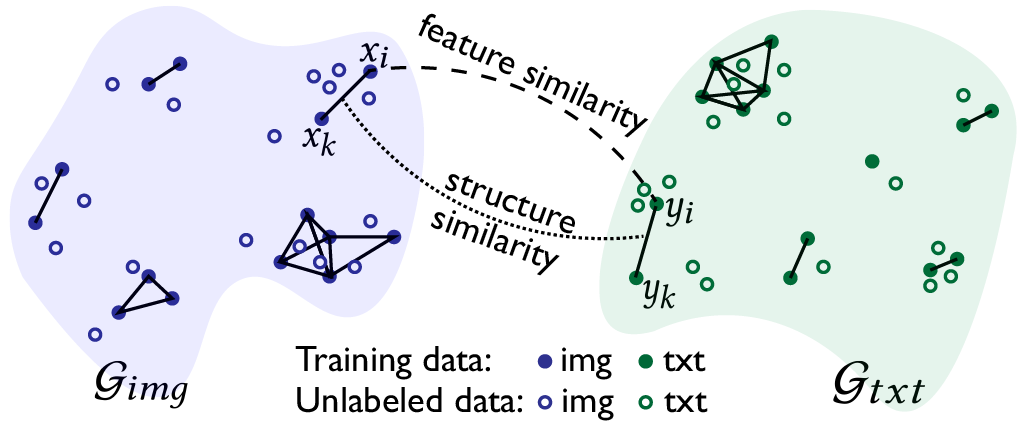}
  \caption{Illustration of $\graph_{img}$ and $\graph_{txt}$  
  and subgraphs $\graph_{txt}^{\mathcal{T}}$, $\graph_{img}^{\mathcal{T}}$ from the labeled training data $\mathcal{T}$.
  Each vertex represents an embedding. Each edge represents high similarity between vertices. Low similarity edges are omitted.}
  \label{fig:twograph}
  \vspace{-0.15in}
\end{figure}

\subsection{Alignment Difficulty (AD)}
\label{subsec:ad}
We now present a formal analysis on the impact of  training data $\mathcal{T}$ on
alignment performance.  Given the complexity of $\graph_{txt}$ and
$\graph_{img}$,  directly modeling or evaluating alignment outcomes is
challenging. Instead, we propose an indirect metric,  {\em Alignment
Difficulty} (AD), to estimate 
the hardness of the alignment task for a given $\mathcal{T}$. 
Our hypothesis is that AD reflects the amount of learning capacity necessary
to learn the new cross-domain knowledge between the two embedding spaces provided by $\mathcal{T}$.
Therefore, the larger the AD, the  harder it is to find a practical
model carrying such learning capacity, and the poorer the alignment
performance.

Following this consideration, we define AD as the distance between two
graphs defined
by the training data $\mathcal{T}$ ($\graph_{txt}^{\mathcal{T}}$ and $\graph_{img}^{\mathcal{T}}$),
which are the subgraphs of $\graph_{txt}$ and $\graph_{img}$ , \ie $\graph_{img}^{\mathcal{T}}
\subset\graph_{img}$, $\graph_{txt}^{\mathcal{T}}
\subset\graph_{txt}$.  As illustrated by Figure~\ref{fig:twograph},  $\graph_{img}^{\mathcal{T}}$ only contains
visual vertices included in $\mathcal{T}$ and
so does $\graph_{txt}^{\mathcal{T}}$.
In this figure,  we illustrate a sample cross-domain similarity binding between a visual embedding $x$ and a textual embedding $y$, reflecting the initial knowledge used by feature-based alignment. 
We also include a sample structure similarity binding between two visual and textual edges, used by structure-based alignment.

We formulate AD as the amount of learning effort required to
{\bf update} the alignment task's initial knowledge 
(defined by $D_{img:txt}(.)$, $D_{img}(.)$ and $D_{txt}(.)$) to match the joint distribution displayed by training data  $\mathcal{T}$. 
Assuming the alignment process considers both feature- and structure-based
principles, we calculate AD as the weighted sum of 
distances between the two subgraphs, $\graph_{txt}^{\mathcal{T}}$ and
$\graph_{img}^{\mathcal{T}}$, under both principles: 
\begin{align}
    & AD({\mathcal{T}}) = 
    \frac{\alpha}{N} \cdot \sum_{(x_i,y_i)\in \mathcal{T}} \; D_{img:txt}(x_i,y_i)   \;\;\; +  \notag \\
       & \frac{1-\alpha}{N^2} \cdot \sum_{(x_i,y_i), (x_k,y_k)\in \mathcal{T}} \; \left|D_{img}(x_i, x_k) - D_{txt}(y_i,y_k)\right|  \label{eq:ad}
\end{align}
Here $\alpha$ ($0\leq \alpha \leq 1$) is an alignment parameter,
representing the weight placed on the feature-based alignment. For clarity, we hereby refer to the unweighted first and second terms in
equation (\ref{eq:ad}) as feature AD and 
structure AD, respectively. 
The three distance metrics, $D_{img:txt}(.)$, $D_{img}(.)$ and
$D_{txt}(.)$, are generic distance metrics cross-domain,
within the visual domain, and within the textual domain, respectively.  
With the goal of making AD model-agnostic, we compute AD assuming $\graph_{txt}$ and
$\graph_{img}$ are constructed from CLIP
embeddings~\cite{radford2021learning} (details in \S\ref{subsec:setup}).

We note that our AD metric  is inspired by the Fused Gromov-Wasserstein (FGW)
distance~\cite{titouan2019optimal, chen2020graph} that 
measures the optimal transport between two structured graphs
{\em in absence of any labeled training data}.  We adapt the formulation of
FGW to calculate the graph distance when the alignment is guided by
the labeled training data $\mathcal{T}$.  This {\em supervised}
setting also reflects the training process of text-to-image
generative models.

\para{Training-from-scratch vs. Fine-tuning.}  It is clear that equation (\ref{eq:ad})  applies to the training-from-scratch scenario. To compute AD when $\mathcal{T}$ is used to fine-tune a (benign) base model, a naive approximation is to mix $\mathcal{T}$  with the training data of the base model $\mathcal{T}_{0}$ and compute $AD(\mathcal{T}\cup \mathcal{T}_0)$.  Since $|\mathcal{T}_{0}|\gg|\mathcal{T}|$, 
fine-tuning data would have negligible effect on $AD(\mathcal{T}\cup
\mathcal{T}_0)$.   Instead, we argue that this approximation is
inaccurate because the impact of fine-tuning data on AD should reflect
their impact on model weights. 
Fine-tuning a model does not start with randomly initialized weights, 
but with those already learned from
$\mathcal{T}_{0}$ and modifies them with
new data $\mathcal{T}$. In the corresponding graph alignment,
this means that 
an existing mapping function ($\theta_{\mathcal{T}_{0}}$) is 
being modified using $\mathcal{T}$.

Assuming $|\mathcal{T}|$ is sufficiently large,  we propose to estimate the AD of
fine-tuning a benign base model  by a weighted sum: 
\begin{equation}
  (1-\lambda) \cdot AD(\mathcal{T}_{0}) + \lambda \cdot AD(\mathcal{T})
  \label{eq:finetune}
\end{equation}
where $0<(1-\lambda)<1$ is a weight memorization factor. 
Given a base model (trained on $\mathcal{T}_{0}$), we can study the impact of  $\mathcal{T}$ that is used
to fine-tune this base model by examining the behavior and trend in $AD(\mathcal{T})$.
We note that this estimation only applies to the scenario of
fine-tuning a benign model with poisoned training data.

%% file: poisonimpact.tex
\secspace
\subsection{Impact of Poisoned Training Data}
\label{subsec:poisonad}
Next, we formally study the impact of poisoning attacks by examining $AD$
of the training data $\mathcal{T}$.  We study the poisoning scenarios described
by prior work~\cite{shan2023prompt}: poisoning a single concept and
poisoning multiple concepts simultaneously.  As defined in \S\ref{sec:threat}, a concept refers to
a common keyword found in prompts that describes the object(s)
in the image, \eg ``bird'', ``cat'', ``city''~\cite{shan2023prompt}.
For all the scenarios examined below,   the training data
$\mathcal{T}$ contains both benign and poisoned data.

\para{Scenario 1: Poisoning a Single Concept.} \\ Let $p$ be the chosen concept to
poison (\eg ``bird''). Let $\benignperc_p$ be the number of data
samples whose prompts contain $p$ in the training dataset $\mathcal{T}$. Let $\poisonperc_p$ be the number of poisoned samples among them, whose textual labels contain $p$ (\eg
``bird'') but are paired with images of the target concept $t$ (\eg ``chandelier''). In this case,
the overall poisoning ratio $\rho$ is $\poisonperc_p/\totaldata$
($\totaldata=|\mathcal{T}|$).  We assume $N$ is large thus $\rho \leq \frac{\benignperc_p}{\totaldata}\ll 1$. For example,
prior work~\cite{shan2023prompt} assumes $\rho\leq 0.01$.

When we replace $\poisonperc_p$ benign samples with poisoned samples, the
maximum change introduced to $AD$ can be estimated by
\begin{equation}
 \alpha \cdot \rho \cdot \Delta_{feature} + 2(1-\alpha) \cdot \rho \cdot \left(\frac{\benignperc_p+\benignperc_t}{\totaldata} - \rho\right) \cdot \Delta_{structure}
\label{eq:single}
\end{equation}
where $\benignperc_t$ ($\ll {\totaldata}$) is the number of training samples of concept $t$, $\Delta_{feature}$  ($\leq 1$) is the
maximum increase in $D_{img:txt}(.)$  a poisoned sample can
introduce beyond its benign version, and $\Delta_{structure}$  ($\leq
1$) is the maximum increase in structure disparity a poisoned sample
can introduce. Since $\benignperc_p+\benignperc_t \ll \totaldata$ and
$\rho<<1$, the second term  
(representing structure AD) is negligible compared to the
first term.
As such, the maximum increase in AD is bounded by a small value $\alpha \cdot
\rho$.  The detailed derivation of (\ref{eq:single}) is shown in Appendix~\ref{app:proof_eq}.

This analysis shows that,  due to the low proportion of
poisoned data in the dataset, poisoning a single concept does not cause noticeable 
changes to AD. However, this does not imply that the poisoning attack
fails. Rather, it indicates that the difficulty in learning the alignment of
$\mathcal{T}$ is similar to that of its benign counterpart.
A model should also achieve similar effectiveness in learning these two datasets.
Therefore, with sufficient poisoned samples of $p$,  the textual
embeddings of $p$ should now align with the visual embeddings of
$t$, while the unpoisoned concepts are not affected.

\begin{conjecture}[{\bf {\em Effectiveness of Poisoning a Single Concept}}] 
    \label{th:single} 
    Poisoning a single concept with very limited poisoned data 
    has little impact on AD. Thus,  the alignment task 
    can learn the joint distribution displayed by the poisoned training data as effectively as it learns the benign version. 
\end{conjecture}
\noindent This explains the results displayed in
Figure~\ref{fig:bird} (\S\ref{subsec:insight1}) and the summary on $O(\mathcal{G}_{poison}(\concept), \concept)$ (\S\ref{subsec:keytakeaways}).
A prompt on the poisoned concept generates misaligned images
(representing its target concept), while prompts on unpoisoned concepts produce correct images.

\para{Scenario 2: Poisoning Multiple Concepts Simultaneously.} \\
Now consider the case where $\poisonc$ concepts are poisoned, and each poisoned concept $p$ has 
$\poisonperc$ poisoned samples in the training data $\mathcal{T}$.  
Now the overall
poisoning ratio becomes $\rho= \frac{\poisonc\cdot
  \poisonperc}{\totaldata}$ and 
grows linearly with $\poisonc$.  For
example, prior work~\cite{shan2023prompt} assumes $\poisonperc=40$ and
$\totaldata=50,000$. Thus 
$\rho=\poisonc \cdot
0.0008$.  

For feature AD, it is easy to show that the increase caused by poisoned samples
is bounded by $\alpha \cdot \rho \cdot \Delta_{feature}$, which scales
linearly with $C_P$.
In parallel, the inclusion of 
$\poisonc \cdot \poisonperc$ poisoned samples across $\poisonc$ concepts could introduce structural changes to the two
graphs $\graph_{txt}^{\mathcal{T}}$ and $\graph_{img}^{\mathcal{T}}$
and thus change structure AD. The change depends on the
composition of poisoned samples among poisoned concepts and their
relationship to all other benign samples.

Without loss of generality, our formal analysis considers the following
simplified setting. Benign samples of different concepts are
well-separated in both visual and textual embedding spaces,
i.e. $D_{img}(x_1,x_2)=1$ if $x_1$, $x_2$ are of different concepts,
and 0 otherwise; $D_{txt}(y_1,y_2)=1$ if $y_1$, $y_2$ are different
concepts, and 0 otherwise.

We prove that when poisoning a fixed number of training samples in each poisoned concept, AD increases with the number of concepts poisoned ($\poisonc$). \vspace{-3pt}
\begin{theorem}[Benefits of Poisoning More Concepts] 
  When benign samples of different concepts are well-separated in both visual and textual embedding spaces,  there exists a configuration of the poisoned training data $\mathcal{T}$ such that AD increases with $\poisonc$, the number of poisoned concepts in $\mathcal{T}$.
\label{th:m} \vspace{-3pt}
\end{theorem}
\noindent The proof of Theorem~\ref{th:m} is in Appendix~\ref{subsec:proof_thm}.

\para{Scenario 3: Model Implosion.} \\
Since AD increases with $\poisonc$, one would ask what happens to the alignment task as $\poisonc$ continues to grow.  As discussed earlier, AD reflects the amount
of learning capacity necessary to  capture the joint distribution displayed by the training
data.  Therefore, we argue that when AD exceeds some value, the alignment task is no longer
feasible in practice. Instead, the alignment will produce a highly erroneous mapping via ``averaging''.  

Here we conjecture that, when $\poisonc$ is sufficiently large, the
poisoned training data $\mathcal{T}$ contains a significant number of
``entangled''  text/image samples whose joint distribution can no
longer be accurately captured and learned by the alignment task.  In
particular, within $\mathcal{T}$, a poisoned concept $p$ is not only
associated with its benign data (\ie visual embeddings of $p$) and
poisoned data (visual embeddings of $t$), but also data from other concepts whose textual or visual embeddings are entangled with those of $p$ or $t$.

Therefore,  as $\poisonc$ increases, the level and scale of the intra-
and inter-graph entanglements also increase.  Beyond a certain
point, each concept $c$, whether poisoned or not, becomes deeply
entangled with multiple other concepts, making the corresponding
joint distributions too complex for the alignment model to
accurately learn.
In this case, the model tends to learn a mapping as some weighted
combinations of the entangled data, often producing incorrect or
``undefined'' visual embeddings with {minimal informative content}.
This is reflected by our empirical observations in
\S\ref{subsec:insight2} where cross-attention maps appear as
``scattered random noise'' (see Figure~\ref{fig:twotrain} and the
observation in \S\ref{subsec:keytakeaways}).  

\begin{conjecture}[Model Implosion] 
  When $\poisonc$ is sufficiently large,  AD of the poisoned
  training data $\mathcal{T}$ exceeds the learning capacity of the alignment model. Thus the learned alignment for a textual embedding becomes a weighted average of the associated visual embeddings introduced by $\mathcal{T}$.  This could produce ``undefined'' visual embeddings with minimal informative content. 
\label{th:implosion} 
\end{conjecture}

\vspace{-0.1in}
\subsection{Limitations of Our Analysis}
\label{subsec:adlimit}
By abstracting the cross-attention mechanism in diffusion models as a task of supervised
graph alignment,  we develop an analytical model to model (and
explain) behaviors of image generative models under poisoning
attacks. However, employing this abstraction also presents several limitations for our work.

\para{Find Exact AD for Model Implosion.}  Our analysis cannot pinpoint a specific threshold on AD, 
beyond which the model implodes.   This threshold depends on many empirical factors, including model
architecture and embedding space configurations.

\para{Compare AD across Tasks.} The absolute value of AD depends on the configurations of the visual and textual feature spaces, as well as the output distributions of multiple distance metrics, which differ across task datasets. Consequently, one should not
directly compare absolute AD values across different task datasets directly.

\para{Fine-tuning Poisoned Models.}  Our equations (\ref{eq:ad}) and
(\ref{eq:finetune}) only apply to  scenarios where a benign base model
is fine-tuned with poisoned training data.  We leave the task of
computing AD for fine-tuing an already poisoned base model to future work.

%% file: setup1.tex
\section{Validating the Analytical Model}
\label{subsec:evalad}

\begin{table*}[t]  
  \resizebox{0.98\textwidth}{!}{%
  \begin{tabular}{c|c|cccc|>{\centering}p{1.2cm}>{\centering}p{1.2cm}>{\centering}p{1.2cm}|>{\centering}p{1.2cm}>{\centering}p{1.2cm}>{\centering}p{1.2cm}|c}
  \toprule
  & $\poisonc$: $\#$ of  & Feature & Structure & AD  & AD  & \multicolumn{3}{c|}{Generation Accuracy} & \multicolumn{3}{c|}{Generation Aesthetics} & Model  \\
  \cline{7-12}
  & Poisoned & AD & AD & ($\alpha=0.8$) & ($\alpha=0.5$)  & All  & Clean  & Poisoned  & All  & Clean  & Poisoned  & Utility\\
  & Concepts& & & & & Concepts & Concepts & Concepts & Concepts & Concepts & Concepts & \\
  \bottomrule
  benign &0   & 0.514 & 0.1493 & 0.441 & 0.331  & 0.90 &  0.90 & -  &0.906 & 0.906 & - & \textbf{0.884}\\
  \bottomrule 
  dirty-label &100 & 0.551 & 0.1475 & 0.470 & 0.349  & 0.654 & 0.656 & 0.638 & 0.849 &0.85 &0.838 & \textbf{0.576}\\
  poison&250 & 0.608 & 0.1463 & 0.515 & 0.377  & 0.433 & 0.435 & 0.427 & 0.817 & 0.815 & 0.824 & \textbf{0.416}\\
  &500 & 0.703 & 0.1440 & 0.592 & 0.424 & 0.357 & 0.34 & 0.374  & 0.766 & 0.761 & 0.771 & \textbf{0.356} \\
    \bottomrule
  clean-label & 100 & 0.520 & 0.1478 & 0.446 & 0.334 & 0.726 & 0.717 & 0.81 & 0.849 & 0.851  & 0.834 & \textbf{0.663}\\
  poison&250 & 0.531 & 0.1445 & 0.453 & 0.338 & 0.62 & 0.625 & 0.604 & 0.782 & 0.781 &0.794 & \textbf{0.552}\\
  & 500 & 0.547 & 0.1369 & 0.465 & 0.342 & 0.566 & 0.548   & 0.584 & 0.721 & 0.727 & 0.715& \textbf{0.473}\\
  \bottomrule
  \end{tabular}
  }
  \caption{AD and model performance, when fine-tuning the {\tt SD1.5} base model using either benign or poisoned training data, under dirty-label and clean-label poisoning attacks.}
  \vspace{-0.1in}
  \label{tab:ad_laion}
\end{table*}

We validate our analytical
model through empirical experiments,  exploring the relationship between AD  (computed
directly from the training data)  and the performance of
trained generative models. We perform
comprehensive experiments by varying datasets, diffusion model
architectures, VAEs, training scenarios
(training-from-scratch/fine-tuning), and poison composition (clean-label/dirty-label).  Besides validation of analytical model,  we also identify
critical and unforeseen findings from these experiments, 
extending beyond those in \S\ref{sec:threat} that motivated our
analytical study. 

\secspace
\subsection{Experimental Setup}
\label{subsec:setup}

\para{Datasets.} Our experiments focus on
LAION-Aesthetics~\cite{laion-aes}, the preferred dataset
for training and studying diffusion models due to its substantial
size and diversity. LAION-Aesthetics includes 120 million high-quality
text/image pairs, covering more than 10,000 object nouns.  We use two secondary 
datasets, CIFAR10~\cite{cifar} and ImageNet~\cite{deng2009imagenet}, to further validate our analysis.  All three datasets are publicly available
with no report of CSAM.

In the following, we describe the experimental setup for LAION-Aesthetics experiments. We configure CIFAR10 and ImageNet
experiments using a similar methodology, with slight 
modifications since they target classification tasks and contain images of lower quality and diversity (details in \S\ref{subsec:cifar+imagenet}).

\para{Model Training.} We consider both fine-tuning and
training-from-scratch scenarios.  For fine-tuning, we consider three 
pretrained base (benign) models:  {\tt SD1.5},  {\tt SD2.1},  and  {\tt
SDXL} models~\cite{rombach2022high}. Each of these high-performance models is trained on more than
170M text/image pairs. We fine-tune
each base model with 50K text/image pairs, which include both benign samples
randomly sampled from LAION-Aesthetics
and poisoned samples (discussed below).  We follow the released training method for Stable Diffusion\footnote{\url{https://github.com/CompVis/stable-diffusion}} and the hyperparameters described by~\cite{rombach2022high}.   
We also train latent diffusion models from scratch with 150K
text/image pairs, following the same data distribution used by our 
fine-tuning experiments.
When training
poisoned LAION-Aesthetics models, we follow~\cite{shan2023prompt} to set the learning rate to 4e-5.

\para{Configuring Poisoned Data.} We adopt a methodology
consistent with prior work~\cite{shan2023prompt}.  We consider poisoning common concepts by randomly selecting from a pool of 500 frequently used nouns that describe objects. 
Here we compute word frequency from 
prompts in
LAION-Aesthetics, and avoid generic
words like ``photo'' or ``picture'' that are present in most prompts.

We consider both dirty-label
and clean-label poisoning attacks. To poison a concept $p$, we
first choose its target concept 
$t$ randomly ($t \neq p$).
We generate a dirty-label poison image $x_t$ by prompting  {\tt SD1.5} with ``a photo of $t$.''  We pair $x_t$ with a text prompt $y_p$ on concept $p$,  extracted from the  LAION-Aesthetics dataset.
Thus $(x_t, y_p)$ is a dirty-label poisoned sample for concept $p$.
Next,  to build clean-labeled poisons,  
we generate clean-label poison images using ~\cite{shan2023prompt}'s released code\footnote{\url{https://github.com/Shawn-Shan/nightshade-release}}. 
We perturb an image $x_p$ from poisoned concept $p$ towards some image
$x_t$ of target $t$ in the visual embedding space. This produces a
perturbed image $x'_p$. We extract a text prompt $y_p$ on concept $p$,
and use $(x'_p,y_p)$ as a clean-label poisoned sample.
The final training data of size $N$ consists of $m=0.0008\cdot N$ poisoned
samples from each poisoned concept and $N-m\cdot C_P$ benign samples randomly selected from LAION-Aesthetics.

\para{Computing AD.} 
Given training data $\mathcal{T}$, we calculate its AD following
equation (\ref{eq:ad}). Since training samples are raw text/image
pairs while AD operates on visual/textual embeddings,  we first apply a
{\em universal} feature extractor $\mathcal{E}(.)$~\cite{radford2021learning}  to convert raw samples into
embeddings.
Let $d_{cos}(.)$ be the cosine distance, $x$ be the image, and $y$ be
the text prompt. We have 
$D_{img:txt}(x,y) = d_{cos}(\mathcal{E}(x),\mathcal{E}(y))$, 
$D_{img}(x_1,x_2)= d_{cos}(\mathcal{E}(x_1),\mathcal{E}(x_2))$,  
$D_{txt}(y_1,y_2)= d_{cos}(\mathcal{E}(y_1),\mathcal{E}(y_2))$. We
normalize each metric into $[0,1]$.

We set $\alpha=0.8$ when computing AD. This choice is informed by prior
work~\cite{chen2020graph} that empirically shows $\alpha=0.8$ to be
superior to other values for computing graph optimal transport.
Existing studies (e.g. ~\cite{livi2013graph}) also suggest that feature-based
alignment holds greater significance than structure-based alignment,
because it captures more meaningful similarities between nodes across
graphs (text vs. image in this case).  
For reference, we also provide AD values for $\alpha=0.5$. The two $\alpha$ values lead to the same conclusion.

\para{Evaluating Generative Models.} To evaluate trained generative
models, we prompt them to generate images and then evaluate these
images. 
We use the default parameters for generation with a guidance scale of $7.5$ and $50$ diffusion steps.
We evaluate LAION-Aesthetics models using prompts on a 
collection of 1000 commonly used concepts. 
This pool consists of the aforementioned pool of 500 nouns to select poisoned
concepts and another pool of 500 frequently used nouns.  The concepts
in these two pools do not overlap, and $78.4\%$ of the concepts in the second pool do
not have any synonyms~\cite{bird2009natural} in the first pool. 
Following existing works~\cite{rombach2022high, ramesh2022hierarchical,ruiz2023dreambooth, park2021benchmark},  we prompt each generative model with 
1000 concepts,  using the prompt of ``a photo of $\concept$'', and collect 5 generated images per concept.  We report the generation performance on 
poisoned and clean (unpoisoned) concepts.

We evaluate the generated images using 
three metrics:
\begin{packed_itemize}
\item {\em {\bf Generation Accuracy}} measures the degree of 
alignment between the generated images and their input prompts; 
\item {\em {\bf Generation Aesthetics}} measures visual
aesthetics of the generated images, \ie harmony and appeal of visual elements that affect perception and interpretation of an image and objects within.
\item {\em {\bf Model Utility}} uses both prompt alignment and visual aesthetics of the generated images to assess model usability.  
A detailed explanation of including this metric can be found in Appendix~\ref{app:justifymetric}.
\end{packed_itemize}

We compute {\em generation accuracy} by studying the statistical distribution
of the CLIP score~\cite{radford2021learning} from generated images and corresponding prompts.
While the CLIP score estimates
the resemblance between an image and a text,  there is no known
threshold for ``accuracy''.  After examining the CLIP
score distribution from a benign generative model ({\tt SD1.5}), we opt to use $0.236$, corresponding to its $10^{th}$ percentile value, as the accuracy threshold. We manually inspect the text/image pairs
used in our experiments to verify it is 
a fair assessment of generation accuracy for these models. For reference, the mean and standard deviation of the CLIP
score for the benign {\tt SD1.5} model is
$0.273\pm0.0276$. 
Figure~\ref{fig:clip_and_aes} and Appendix~\ref{app:justifymetric} illustrate the effectiveness of this accuracy measure.  

We compute the visual quality of the generated
images using the CLIP aesthetics score~\cite{laion-aes} and 
apply the threshold of $6.5$ as suggested by~\cite{laion-aes}.  Recent
works have validated the use of CLIP aesthetics to assess image
visual quality~\cite{hentschel2022clip,wang2023exploring,xu2023clip}, whereas
alternative measures such as fr\'echet inception distance (FID) are
shown to be ineffective in measuring the visual quality of generated
images~\cite{podell2023sdxl,jayasumana2023rethinking,kynkaanniemi2022role}.

With these in mind, the exact performance metrics are as follows: 
\begin{packed_itemize} 
\item {\em \textbf{Generation Accuracy}} : \% of 
generated images with CLIP$> 0.236$, 

\item {\em \textbf{Generation Aesthetics}} : \% of images with aesthetics$> 6.5$, 

\item  {\em \textbf{Model Utility}} : \% of images with CLIP$>
0.236$ and aesthetics$> 6.5$. 
\end{packed_itemize}

%% file: verification1.tex
\secspace
\subsection{Main Results of LAION-Aesthetics Experiments}
\label{subsec:mainlaion}
In this section, we report the key results from {\bf fine-tuing the pretrained  {\tt SD1.5} model}.  We discuss the results of  fine-tuning other base models and training models from scratch later in \S\ref{subsection:ablation}. 

We experiment with both single-concept poisoning ($\poisonc = 1$) and many-concept poisoning ($\poisonc\geq$100).  Results on single-concept poisoning are as expected: they barely changed AD, successfully manipulated image generation on poisoned concepts, but had minimal influence on unpoisoned concepts. The detailed result can be found in Appendix~\ref{app:fewconcept}.

We now focus on many-concept poisoning. We consider a very weak poisoning scenario, by limiting the number of poisoned training samples per poisoned concept ($m_p$) to 40. The overall poison ratio on the fine-tuning data $\mathcal{T}$ is $\poisonc \cdot 0.08$\%.  We experiment with both dirty-label and clean-label attacks and list their results separately.  As expected, since they produce different types of poisoned data, their AD values are different.

\begin{figure}[t]
	\centering \vspace{-0.15in}
  \includegraphics[width=0.32\textwidth]{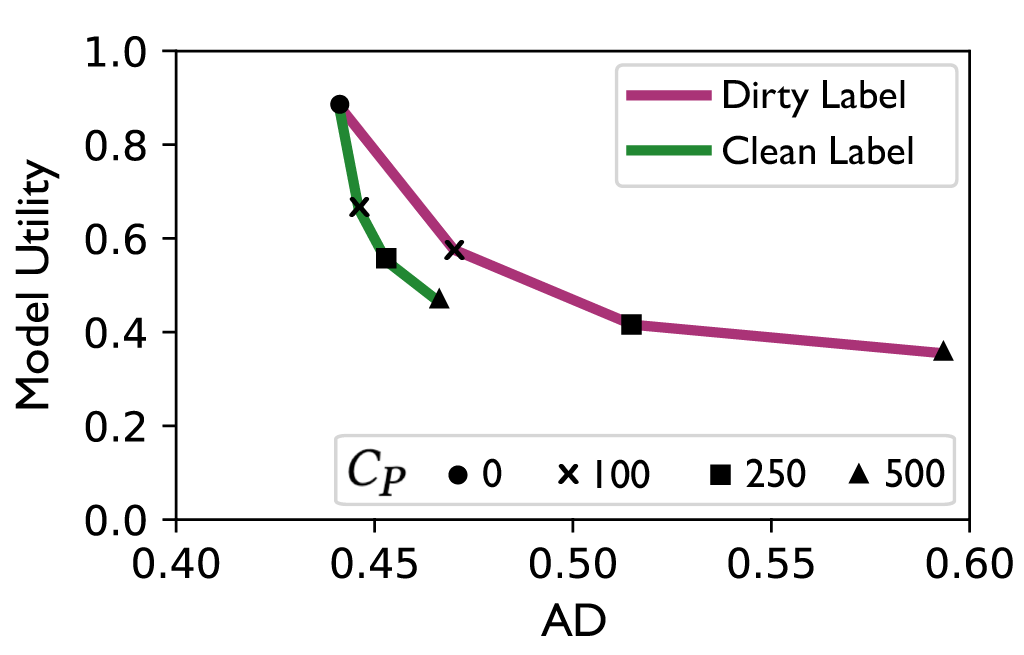} \vspace{-0.1in}
  \caption{Strong correlation between AD and generation performance of fine-tuned {\tt SD1.5} models, for both dirty-label and clean-label poisoning attacks.}
  \label{fig:ad_perf_finetune}
  \vspace{-0.in}
\end{figure}

Table~\ref{tab:ad_laion} presents the overall result, in terms of AD and detailed model performance under different training data configurations.  We make the following key observations, which apply to both dirty- and clean-label poisoning attacks. 
\begin{packed_itemize} \vspace{-1pt}
\item AD increases with the number of poisoned concepts ($\poisonc$), dominated by the increase in feature AD.
\item The model performance declines with $\poisonc$, for both poisoned and clean (unpoisoned) concepts.   
\item There is a strong connection between AD and the performance of the poisoned generative model, better illustrated by  Figure~\ref{fig:ad_perf_finetune}. 
\item  The effect of model implosion is evident\footnote{As an additional verification, we manually inspect the cross-attention map for the generated images, and verify that the majority of them follow the ``scattered random noise'' pattern seen in Figure~\ref{fig:twotrain}.} and its severity intensifies as $\poisonc$ increases.
  \vspace{-1pt}
\end{packed_itemize}

We make further observations on dirty and clean-label attacks. 

\para{Dirty-Label Attacks.}  As $C_P$ increases, the feature AD displays a strong increase, since dirty-label samples generally display larger $D_{img:txt}(.)$ values.  Yet the impact on structure AD is less visible and displays a weak, decreasing trend with $\poisonc$.  We think this comes from two factors. First, since both the poisoned concepts and their targets are randomly chosen, the poisoned data introduces structural changes in various uncoordinated directions, which can either amplify or cancel each other's effects, resulting in a less noticeable impact on structure AD.  Second, the structural properties of benign training samples are already complex. The poisoned data introduces different but relatively stronger associations between text prompts and images, which slightly reduces structure AD.   Nevertheless, as feature AD consistently outweighs structure AD,  the overall AD continues to grow as $\poisonc$ increases.

\para{Clean-Label Attacks.}  Another key observation is that concurrent clean-label poisoning attacks can lead to model implosion, but at a slower pace (i.e., it requires poisoning more concepts) compared to its dirty-label counterpart.   This can be explained by its AD value. For the same $\poisonc$ value, AD of clean-label poisoned data is lower than that of dirty-label poisons (see Figure~\ref{fig:ad_perf_finetune}). This is as expected because these perturbed images carry smaller $D_{img:txt}$. On the other hand, this property also makes them stealthy and hard to detect as they ``blend'' into benign samples. 

We note that this finding also suggests the potential of improving the design of stealthy, clean-label attacks, including selecting  target concepts and the images to perturb,  so that their attack potency can further approach that of dirty-label attacks.  We leave this to future work.

\para{Poisoning Less Popular Concepts.}   So far, our experiments have selected poisoned concepts from the top 500 frequently used concepts.  We also examine the poison effect of poisoning less popular concepts. Specifically, we find 500 concepts whose word frequency ranking is between 1500 and 2000, and produce dirty-label poisoned samples as before. Each target concept is also randomly chosen from the same pool. We then evaluate the trained model using 500 concepts whose word frequency ranks between 501 and 1000,  and are considered benign (unpoisoned) for both models.  In this case, the training data's AD is 0.599 (nearly identical to that of poisoning top 500 concepts), and so is the model utility on these benign concepts (0.375 vs. 0.392).   This further demonstrates the generality of model implosion and the strong tie between AD and model performance.

\begin{table}[t]
  \resizebox{0.45\textwidth}{!}{
  \begin{tabular}{c|c|c}
  \toprule
  Word Frequency Ranks of & AD & Model Utility\\
  $C_P$=500 Poisoned Concepts& ($\alpha = 0.8$) & (500 Benign Concepts) \\
  \midrule
  1-500 & 0.592 & \textbf{0.375} \\
  \midrule
  1501-2000 & 0.599 & \textbf{0.392}\\
  \bottomrule
  \end{tabular}
  }
  \caption{AD and model utility of two fine-tuned {\tt SD1.5} models,  by poisoning top 500 frequently used concepts (1-500), or those ranked 1501-2000.  We test the models with 500 unpoisoned concepts ranked 501-1000 in frequency.
  }     \vspace{-0.2in}
  \label{tab:ad_lowfreq}
\end{table}

\begin{figure}[t]
    \centering
    \includegraphics[width=0.32\textwidth]{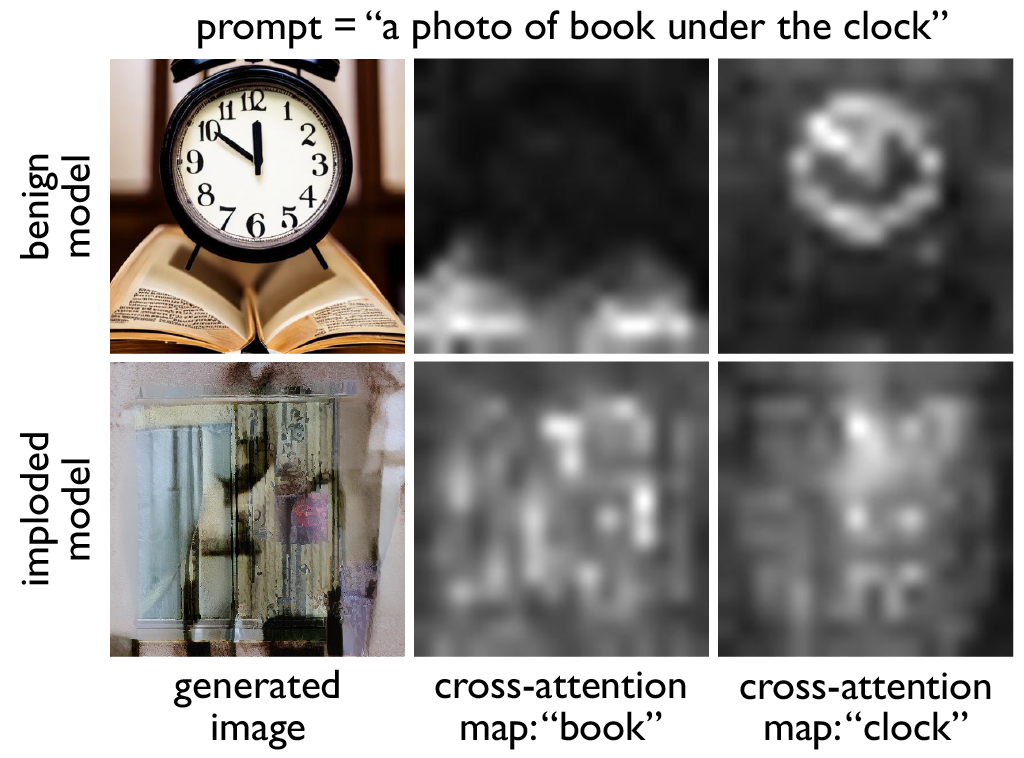} \\\vspace{0.05in}
    \includegraphics[width=0.32\textwidth]{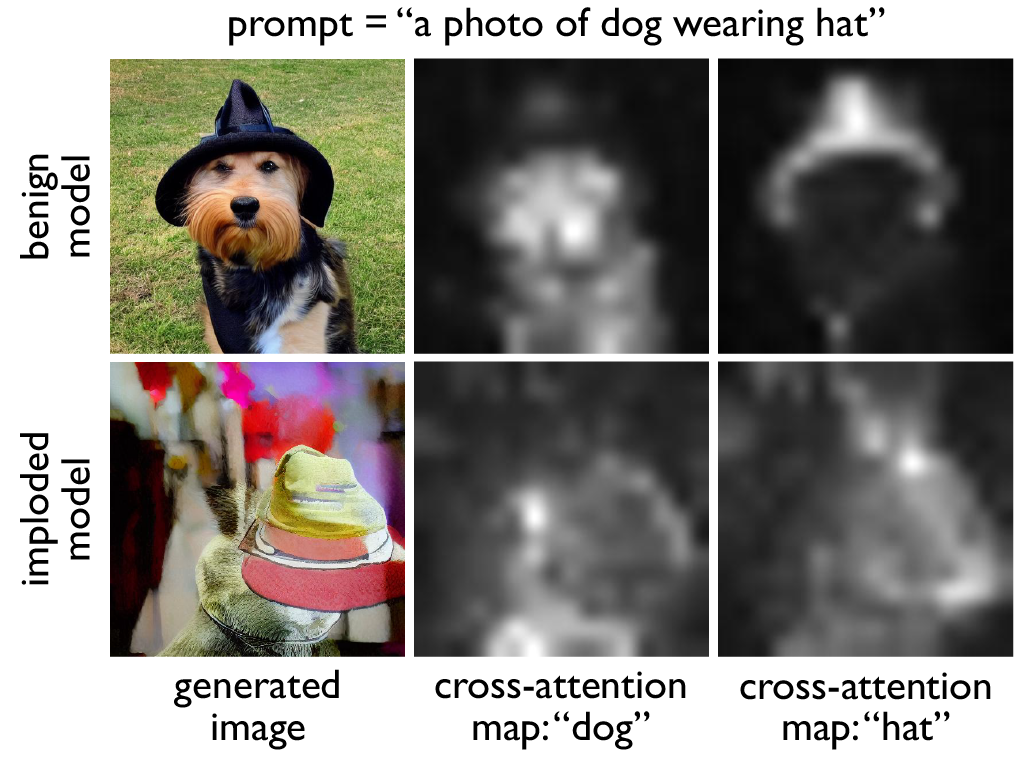} \vspace{-0.1in}
    \caption{Generated images and token-specific cross-attention maps when prompted
      with multiple concepts.}           
    \label{fig:multi}
    \vspace{-0.1in}
  \end{figure}

\para{Impact on More Complex Prompts.}  We also test the trained generative models using input prompts containing multiple concepts, \eg ``a photo of book under the clock'' and ``a photo of dog wearing hat.''   Here we study multiple cross-attention maps, one per  concept in the input prompt. As shown in Figure~\ref{fig:multi}, when a model implodes, the token-specific cross-attention maps (targeting different concepts) become scattered noise. Moreover, these token-specific maps are highly similar,  indicating that the cross-attention module can no longer distinguish individual concepts in the prompt.  This suggests that model implosion can scale to complicated prompts.  

\secspace
\subsection{Additional LAION-Aesthetics Experiments}
\label{subsection:ablation}
We also perform ablation studies by varying training scenarios and base {\tt SD} models used for fine-tuning.

\begin{table*}[t]  \vspace{-0.0in}
  \resizebox{0.78\textwidth}{!}{%
  \begin{tabular}{c|c|c|>{\centering}p{1.2cm}>{\centering}p{1.2cm}>{\centering}p{1.2cm}|>{\centering}p{1.2cm}>{\centering}p{1.2cm}>{\centering}p{1.2cm}|c}
  \toprule
  $\poisonc$: $\#$ of  & Base & AD & \multicolumn{3}{c|}{Generation Accuracy } & \multicolumn{3}{c|}{Generation Aesthetics} & Model  \\
  \cline{4-9}
  Poisoned & Model & ($\alpha=0.8$) & All & Clean & Poisoned & All & Clean & Poisoned  & Utility \\
  Concepts & & & Concepts & Concepts & Concepts & Concepts & Concepts & Concepts  &   \\
  \midrule
0 & {\tt SD2.1} & 0.441 & 0.912 & 0.912 & - & 0.950 & 0.950 & - & \textbf{0.891} \\
\bottomrule 
100 & {\tt SD2.1} & 0.470  & 0.854 & 0.854 & 0.850 & 0.288 & 0.288 & 0.290 & \textbf{0.181}\\
250 & {\tt SD2.1} & 0.515 & 0.814 & 0.827 & 0.776 & 0.279 & 0.282 & 0.269 & \textbf{0.170}\\
500 & {\tt SD2.1} & 0.592 & 0.715 & 0.740 & 0.690 & 0.286 & 0.307 & 0.266 & \textbf{0.116}\\
\midrule
0 & {\tt SDXL} & 0.441 & 0.882 & 0.882 & -  & 0.983 & 0.983 & - & \textbf{0.877} \\
\bottomrule 
100 & {\tt SDXL} & 0.470 & 0.564 & 0.565 & 0.558 & 0.744 & 0.741 & 0.774 & \textbf{0.495} \\
250 & {\tt SDXL} & 0.515 & 0.519 & 0.520 & 0.516 & 0.738 & 0.737 & 0.742 & \textbf{0.456} \\
500 & {\tt SDXL} & 0.592 & 0.473 & 0.474 & 0.472 & 0.656 & 0.654 & 0.658 & \textbf{0.384} \\
  \bottomrule                    
  \end{tabular}
  }
  \caption{Performance of dirty-label poisoning attacks on two additional base models, {\tt SD2.1} and {\tt SDXL}. }
  \vspace{-0.13in}
  \label{tab:ad_change_archi}
\end{table*}

\begin{table*}[t]  
  \resizebox{0.85\textwidth}{!}{%
  \begin{tabular}{c|ccc|>{\centering}p{1.2cm}>{\centering}p{1.2cm}>{\centering}p{1.2cm}|>{\centering}p{1.2cm}>{\centering}p{1.2cm}>{\centering}p{1.2cm}|c}
  \toprule
  $\poisonc$: $\#$ of  & Feature & Structure & AD & \multicolumn{3}{c|}{Generation Accuracy} & \multicolumn{3}{c|}{Generation Aesthetics} & Model  \\
  \cline{5-10}
  Poisoned & AD & AD & ($\alpha=0.8$) & All  & Clean  & Poisoned  & All  & Clean  & Poisoned  & Utility \\
  Concepts& & &  & Concepts & Concepts & Concepts & Concepts & Concepts & Concepts & \\
  \midrule
  0 & 0.513 & 0.1490 & 0.440 & 0.762 & 0.762 & - & 0.894 & 0.894 & - & \textbf{0.728}\\
  \bottomrule 
  100 & 0.551 & 0.1472 & 0.470   & 0.637 & 0.643 & 0.580 & 0.693 & 0.694 & 0.680 & \textbf{0.522}\\
  250 & 0.608 & 0.1462 & 0.516   & 0.540 & 0.537 & 0.548 & 0.752 & 0.752 & 0.752 & \textbf{0.468}\\
  500 & 0.703 & 0.1437 & 0.592   & 0.518 & 0.516 & 0.52  & 0.772 & 0.771 & 0.773 & \textbf{0.478} \\
  \bottomrule                    
  \end{tabular}
  }
  \caption{AD and model performance, when training latent diffusion models from scratch, under dirty-label poisoning attacks.}\vspace{-0.1in}
  \label{tab:ad_laion_scratch}
\end{table*}

\para{Varying Base Diffusion Model.} Besides {\tt SD1.5}, we also experiment with fine-tuning {\tt SD2.1} and {\tt SDXL}~\cite{stable2, stable2.1, podell2023sdxl} models with poisoned training data.  As discussed in \S\ref{subsec:back_diffusion}, {\tt SD2.1} uses the same VAE  as  {\tt SD1.5} but a different decoder fine-tuned on high-quality images,  while {\tt SDXL} retrains the VAE on new, high-quality images~\cite{podell2023sdxl}.  Since our empirical calculation of AD uses the same universal encoder,  the AD values for both are the same as those for {\tt SD1.5} when using the same training data. 

Results in Table~\ref{tab:ad_change_archi} show that poison is more effective against {\tt SD2.1} and {\tt SDXL} models,  leading to a faster degradation in model utility and a faster pace into model implosion.  Interestingly, for poisoned {\tt SD2.1} models, the generation aesthetics drops dramatically to below 30\%.   A closer look reveals that many generated images are just a single color block (visual examples are shown in Appendix~\ref{app:scratch_and_archi}).
This also confirms that a metric combining both CLIP score and aesthetics, \ie model utility, is more reliable. 

We suspect this fast degradation is because {\tt SD2.1} and {\tt SDXL} 's VAE encoder and/or decoder are trained on higher quality images, leading to overfitting or lack of generalizable interpretation on the visual embeddings~\cite{stable2,podell2023sdxl}.  Thus when an imploding model produces undefined embeddings, the decoder is unable to produce any useful content.  Such instability issues of  {\tt SD2.1} and {\tt SDXL} models are already documented by practitioners~\cite{vass2023explaining,romero2022stable} and {\em \textbf{can further amplify the damage of model implosion.}} 

\para{Training-from-Scratch.} Using 150K training samples constructed from LAION-Aesthetics, we train latent diffusion models ``from scratch''. We use the same method to curate poisoned samples and have 40 poisoned samples in  each poisoned concept.  Table~\ref{tab:ad_laion_scratch} demonstrates the same trends as those observed on fine-tuning  {\tt SD1.5}: (1) AD increases with $\poisonc$,  (2) model performance degrades with $\poisonc$ and there is a strong correlation between AD and model performance, and (3) the model already shows initial symptoms of implosion when poisoning $100$ concepts.

We note that AD values computed on the larger 150K training data are nearly identical to those in Table~\ref{tab:ad_laion}.  This is because the two datasets share almost identical joint distribution under the same  poison ratio and AD is normalized by the training data size $N$.

%% file: cifarresult.tex
\secspace 
\subsection{Experiments on CIFAR10 and ImageNet}
\label{subsec:cifar+imagenet}
We also use two secondary datasets: CIFAR10
and ImageNet, commonly used by image classification tasks.
Compared to LAION-Aesthetics,  these datasets have fewer images and of
lower quality.  Furthermore, the size and diversity of their class labels are
very limited, with CIFAR10 having only 10 classes and ImageNet having
1,000 distinct labels (\eg specific animal breeds rather than common
object names).  Given their limited size and lack of diversity,  we  only use these two datasets to confirm LAION-Aesthetics results.

\para{Experimental Setup.} 
For CIFAR10, we leverage its training set of 50K images (of size
32$\times$32) to study the
scenario of training-from-scratch. We use the latent diffusion
architecture from~\cite{rombach2022high} that matches its image size. For each image, we treat its
class label as the text prompt (since each small image is dominated by the
object described by the class label).  We train poisoned generative models from scratch by
randomly select $\poisonc$ classes to poison.  When poisoning a class, we
include 3000 benign samples with mislabeled text/image pairs. Thus, each poisoned class' training data includes 5000 benign
and 3000 poisoned samples, while a benign class has 5000 benign
samples. Since CIFAR10's  prompt set is too limited (\ie only 10
nouns), we modify AD computation to use the following distance metrics:  
$D_{img:txt}(x,y) = 0$ if $x$ is correctly labeled by $y$, and $1$
otherwise; $D_{img}(x_1,x_2)= d_{cos}(\mathcal{E}(x_1),
\mathcal{E}(x_2))$; $D_{txt}(y_1,y_2)=0$ if $y_1=y_2$, and $1$
otherwise.

The ImageNet's training data includes 1.2M images of size
224$\times$224.  Using the latent diffusion
architecture from~\cite{rombach2022high} matching its image size, we first train a benign
generative model from scratch using the full training set.  For each
image, we use BLIP conditioned on its class label to
produce its text prompt.  We then construct benign and poisoned training data to fine-tune the benign
base model.   The fine-tuning dataset includes $N$=5K samples covering
100 randomly selected classes.  Each poisoned class has $m_p$=50 mislabeled
text/image pairs and each benign class has 50 benign text/image
pairs randomly sampled from the training set.  Consistent with the
LAION-Aesthetics experiments, we randomly select
the poisoned classes and their target classes, and vary the number of poisoned classes ($C_P$) from 10 to 100. 

\begin{table}[t]
\resizebox{0.5\textwidth}{!}{%
  \begin{tabular}{c|ccc|>{\centering}p{1.3cm}>{\centering}p{1.3cm}>{\centering\arraybackslash}p{1.3cm}}
  \toprule
  $\poisonc$: $\#$ of    & Feature  & Structure  & AD  &   \multicolumn{3}{c}{Generation Accuracy} \\
  \cline{5-7}
  Poisoned & AD & AD & ($\alpha=0.8$) &  All & Clean & Poisoned \\
  Classes& & &  &  Classes  &  Classes&  Classes\\ 
  \midrule
  \multicolumn{7}{c}{CIFAR10} \\
  \midrule
  0    &  0      & 0.690   & 0.138   &  0.964 & 0.964 & -  \\ 
  \midrule
  2    & 0.107   & 0.621   & 0.210   &  0.657   & 0.620 & 0.803 \\
  4    & 0.194   & 0.637   & 0.282   & 0.60   & 0.679 & 0.471  \\
  6    &  0.265  & 0.649   & 0.342   & 0.473 & 0.680 & 0.335    \\
  8    & 0.324   & 0.658   & 0.391   & 0.452  & 0.874  & 0.414 \\  
  10   & 0.375   & 0.663   & 0.433   & 0.434 & - & 0.434\\
  \midrule
  \multicolumn{7}{c}{ImageNet} \\
  \midrule
  0    &  0.411   & 0.126   & 0.354   & 0.735  & 0.735  & - \\
  \midrule
  10   &  0.465   & 0.127   & 0.397  & 0.717  & 0.724  & 0.013 \\
  50   &  0.679   & 0.129   & 0.569  & 0.539  & 0.567  & 0.002 \\
  100  &  0.935   & 0.126   & 0.773  & 0.171  & 0.190  & 0.000 \\
 \bottomrule
  \end{tabular}
  }
  \caption{AD and image generation accuracy of models trained on CIFAR10 and ImageNet.}   
  \label{tab:ad_cifar10}
  \vspace{-0.3in}
\end{table}

\para{Model Evaluation.} Due to the limited image size and prompt
space, we evaluate the generative models by the {\em generation accuracy} computed from a classifier trained to
determine the class label of each generated image. 
For CIFAR10, we take a Resnet18
model pretrained on ImageNet and fine-tune the model using its benign
training data.  We test each generative model by generating 1000
images per class and
compute {\em \textbf{generation accuracy}} as \% of generated
images whose classification label matches its prompt. 
For ImageNet, we take the pretrained ResNet50 model and test each model by generating 100 
images for each of the 1000 classes, and report their classification accuracy.

\para{Results.} Table~\ref{tab:ad_cifar10} lists the AD values 
and the classifier-based generation accuracy for CIFAR10 and ImageNet models.
We include results for poisoned and clean
(unpoisoned) classes separately.  Consistent with the
LAION-Aesthetics experiments,  we observe a strong correlation between AD and model generation accuracy.

We observe some minor differences among the results of the
LAION-Aesthetics, CIFAR10, and ImageNet experiments, likely due to
their inherent disparities
in data characteristics. For CIFAR10
(Table~\ref{tab:ad_cifar10}),  the generation performance of benign
classes is much less affected by the growing number of poisoned
classes ($C_P$). This indicates that when poisoning attacks are
effective, they do not cause the model to implode. This is likely
because the text/image associations in CIFAR-10's 10-class
dataset -- whether benign or poisoned -- are relatively straightforward and
limited in textual diversity. Consequently, a standard latent
diffusion model with cross-attention layers can effectively learn
these associations.  On the other hand, the generation
accuracy, measured by the label classification accuracy, can be volatile
across the 10 classes. This is because these 10 classes are known to display large
inter-class discrepancy for accuracy and robustness~\cite{tian2021analysis, benz2021robustness}. 

\begin{figure}[t]
  \centering
  \vspace{-0.05in}
  \includegraphics[width=0.38\textwidth]{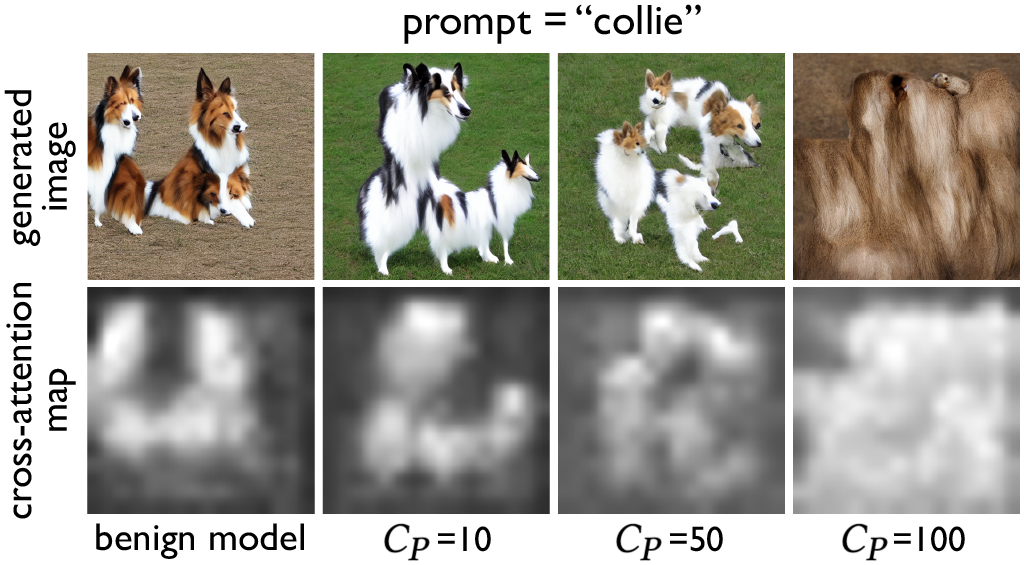} \\\vspace{0.05in}
  \includegraphics[width=0.38\textwidth]{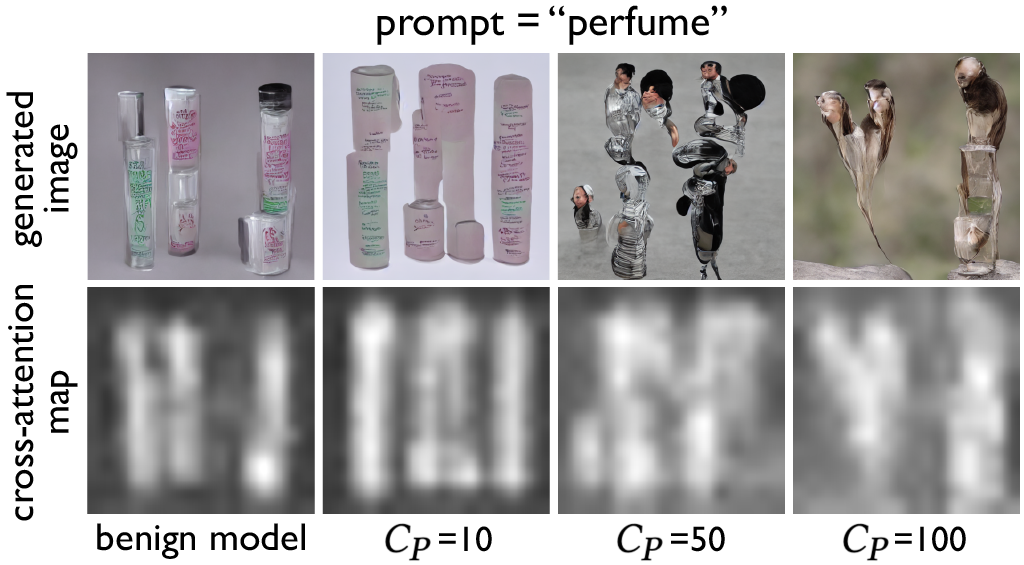} 
  \vspace{-0.1in}
  \caption{Generated images and their token-specific cross-attention
      maps for unpoisoned classes in ImageNet.}         
  \label{fig:imagenet_gen}
  \vspace{-0.1in}
\end{figure}

For ImageNet (Table~\ref{tab:ad_cifar10}),  where the prompt space is
100 times larger than that of CIFAR10, we observe the phenomenon of
model implosion when $C_P$ goes beyond 50.  We further confirm this
observation by studying the generated images and their token-based
cross-attention maps. Figure~\ref{fig:imagenet_gen} plots examples of
generated images for two unpoisoned classes, ``collie'' and
``perfume'',  and their token-based cross-attention maps,  as we
increase $C_P$.  Similar to our LAION-Aesthetics experiments, as more
classes get poisoned,  the training data introduces more complexity
to the joint distribution to be learned. Eventually, the model
implodes, and the unpoisoned classes are significantly affected.

Compared to the LAION-Aesthetics experiments, the poisoned classes in
the ImageNet experiments show significantly lower generation
accuracy, because the classification accuracy metric amplifies the  
impact on generated images that misalign with their text
prompts.
Here we do not use CLIP and aesthetic scores for evaluating ImageNet models because these
metrics are designed based on LAION images and could introduce bias and errors in ImageNet experiments due to the difference in image size and distribution.

%% file: counter.tex
\section{Analysis on Poisoning Defenses} 
\label{sec:counter}

In this section, we leverage our analytical
framework to study the efficacy of potential defenses against poisoning attacks. We explore
three different defense mechanisms from~\cite{shan2023prompt}:  (1) applying image-text alignment
filtering to remove poisoned training samples, (2) filtering out high
loss training data, and (3) fine-tuning the imploded model with only
benign training data.  For all these defenses, we consider clean-label
poisoning attacks. 

\para{Defense 1: Image-Text Alignment Filtering.} Alignment measure
has been used to detect poisoned data~\cite{yang2023data} and 
noisy inputs~\cite{changpinyo2021conceptual, schuhmann2021laion,schuhmann2022laion} in generative models.
This defense uses the
alignment (or similarity) score of each text/image pair to identify and remove 
potentially poisoned samples from the training dataset.   
The hypothesis is that
poisoned data's  image $x$ and text $y$ are less aligned than benign
data. Thus, the model trainer could attempt to remove poisoned data by filtering out those with low
alignment scores.

\begin{figure}[t]
  \centering
  \subfigure[CLIP Score]{
    \includegraphics[width=0.32\textwidth, trim={0 1.1cm 0 0},clip]{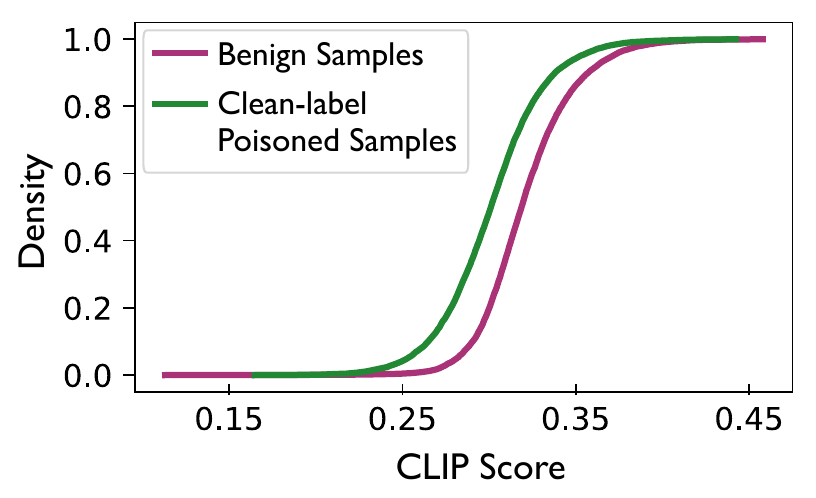}  
    \label{fig:clipscorecdf}
  }
  \subfigure[$D_s(x,y)$]{
    \includegraphics[width=0.32\textwidth, trim={0 1.1cm 0 0},clip]{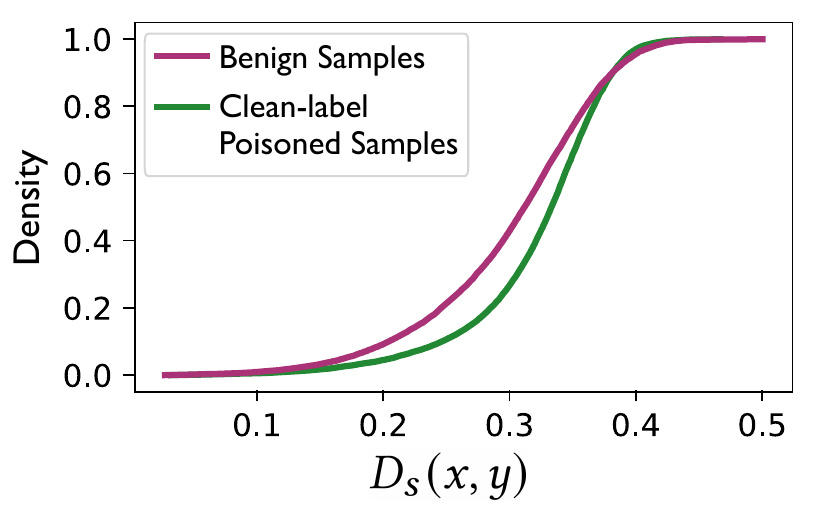} 
    \label{fig:struc_ad_cdf}
  }
  \vspace{-0.15in}
  \caption{Cumulative distribution of (a) CLIP alignment score, (b) $D_s(x,y)$, across benign and poisoned
  training samples used in LAION-Aesthetics experiments.}
  \vspace{-0.15in}
\end{figure}

Under our analytical model, this defense can apply the
alignment score as $D_{img:txt}(x,y)$ in the AD computation, since removing data samples with high $D_{img:txt}(x,y)$ could reduce AD.  We empirically study this
filtering defense by
studying the distribution of $1-D_{img:txt}(x,y)$, calculated from the CLIP score (see \S\ref{subsec:evalad}).  
Figure~\ref{fig:clipscorecdf} plots the cumulative
distribution function (CDF) of the CLIP score for both benign and
poisoned training samples used in our LAION-Aesthetics experiments.   We see that
since the two distributions are similar,  benign and
poisoned data 
exhibit comparable sensitivity to score-based filtering. 
For example, filtering out the lowest 50\%
of poisoned data will remove 21\% of benign data. This explains why
such defense leads to limited impact, empirically verified
by~\cite{shan2023prompt}.

This analysis also suggests that attackers can try to curate
more stealthy 
clean-label poisoned samples (\eg selecting the ``right'' image to perturb and/or optimizing
their text captions~\cite{shan2023prompt})  
without compromising attack potency.  In parallel,  the model trainer can try to identify and apply 
significantly different score functions (or classifiers) that make the score
distributions of benign and poisoned data more distinct.  Yet this
approach still faces the same  challenge that filtering out
poisoned data may remove useful benign data.

\para{Defense 2: Filtering High Loss Data.}  Like the above, another filtering-based defense is to identify 
(and remove) poisoned data that causes high model training
loss. Under our analytical model,  we argue that these high-loss
samples are those $(x,y)$
pairs with high individual $AD(x,y)$ values, defined as
\begin{align}
  AD(x,y)  
  = \alpha \cdot D_{img:txt}(x,y)   +  (1-\alpha) \cdot D_{s}(x,y)  
\end{align}
where $D_{s}(x,y) = \frac{1}{N} \cdot \sum_{(x_k,y_k)\in \mathcal{T}} \; \left|D_{img}(x,x_k) -  D_{txt}(y,y_k)\right|$.
Intuitively, a poisoned sample $(x,y)$ with higher $AD(x,y)$ value
carries more crucial information to be learned, leading to larger
loss during alignment learning.  Note that the first term
$D_{img:txt}(x,y)$ is the same as the above 
defense using 
CLIP alignment scores.  The second term 
$D_s(x, y)$
reflects the structure AD contributed by a single sample
$(x,y)$.

For our LAION-Aesthetics  experiments (\S\ref{subsec:evalad})  we find
that benign and poisoned data display nearly identical distributions
on 
$D_s(x,y)$ (Figure~\ref{fig:struc_ad_cdf}).
This indicates that filtering high
loss data is ineffective regardless of whether it uses $AD(x,y)$, 
$D_{img:txt}(x,y)$, or 
$D_s(x,y)$
to compute loss.   This aligns with the empirical results
in~\cite{shan2023prompt}.

\begin{table}[t]
    \centering
  \resizebox{0.35\textwidth}{!}{
      \begin{tabular}{c|>{\centering}p{1.3cm}>{\centering}p{1.3cm}>{\centering\arraybackslash}p{1.3cm}}
      \toprule
      $\#$ of   & \multicolumn{3}{c}{Generation Accuracy} \\
      \cline{2-4}
      Benign Data & All  & Clean  & Poisoned \\
      in Fine-tuning & Concepts & Concepts & Concepts\\
      \midrule
    Poisoned Model    &  0.566      & 0.548 & 0.584 \\ \midrule
    5K    & 0.660  & 0.652   & 0.668   \\
    10K    & 0.702   & 0.692   & 0.712    \\
    20K    &  0.715  & 0.734   & 0.696   \\
    30K    & 0.733   & 0.718   & 0.748 \\ \midrule
    Benign Model & 0.90  & 0.90  & -  \\
    \bottomrule
    \end{tabular}
  }
    \caption{Generation accuracy after fine-tuning an imploded model
      with benign training data.}
      \vspace{-0.25in}
    \label{tab:fintune_cleanlabel}
  \end{table}

\para{Defense 3: Subsequent Fine-tuning with Benign Data.} When a model implodes,  a natural defense is to
subsequently fine-tune the poisoned model on benign data.   Here  the fine-tuning process uses benign data to update the
``entangled knowledge''
learned by the model and hopes to remove the poison effect eventually. 

Under this scenario,  AD computation 
differs from Equation (\ref{eq:ad}), because the initial cross-attention knowledge (\ie those carried by the
imploded model) is different
from that captured by Equation (\ref{eq:ad}).   We leave the task of
computing AD in this scenario to future work.
On the other hand, by starting from entangled/distorted 
knowledge, the amount of learning on alignment required by fine-tuning an imploded model with benign data
should be notably higher than that of  fine-tuning a benign model with poisoned
data.   As such, it should take more fine-tuning effort and training data to bring an imploded model back to its benign state. 

We verify this hypothesis empirically by fine-tuning an imploded {\tt SD1.5}
model. Recall that this imploded model was first fine-tuned with 50K samples, 30K of which are benign samples. 
Table~\ref{tab:fintune_cleanlabel} lists the model's
generation accuracy after being fine-tuned with 5K to 30K
benign samples. We observe a ``diminishing
return'' effect after 10K samples. Even with 30K benign training
data, the generative model's performance is still far from the
original benign version. Therefore, it is likely more efficient to
revert the model to its latest benign version recorded.

%% file: conclusion.tex
\secspace
\section{Conclusion} 
Our work establishes the first analytical framework to model and
study the impact of data poisoning attacks against large-scale text-to-image generative
models. By abstracting the cross-attention mechanism in generative
models as supervised graph alignment, we formally analyze the
impact of poisoned training data.  
Under this framework, we identify the 
impact of concurrent poisoning attacks on model behaviors, which differs from those of individual attacks. 
This confirms and explains the surprising phenomenon of ``model
implosion''.  We validate our analytical framework with extensive
experiments and identify fresh insights on model implosion. We further apply
this framework to evaluate the efficacy of existing poison defenses.

Moving forward, we plan to leverage this framework to identify more
potent poisoning attacks against diffusion models and their
defenses,
One potential direction is to design 
source-target selection to improve AD. However, these
strategies must also resist defenses that filter out poison samples by
detecting significant 
 AD contributions or reverse-engineering source-target mappings.
 Another direction is to expand our analysis to include a wider range of training 
 scenarios, including fine-tuning a poisoned model. 

\secspace
\subsection*{Acknowledgements}
This work is
supported in part by NSF grants CNS-2241303 and CNS-1949650. Opinions, findings,
and conclusions or recommendations expressed in this material are those of
the authors and do not necessarily reflect the views of any funding agencies.

%% file: appendix.tex
\begin{figure*}[!htb]
  \centering
  \includegraphics[width=0.98\textwidth]{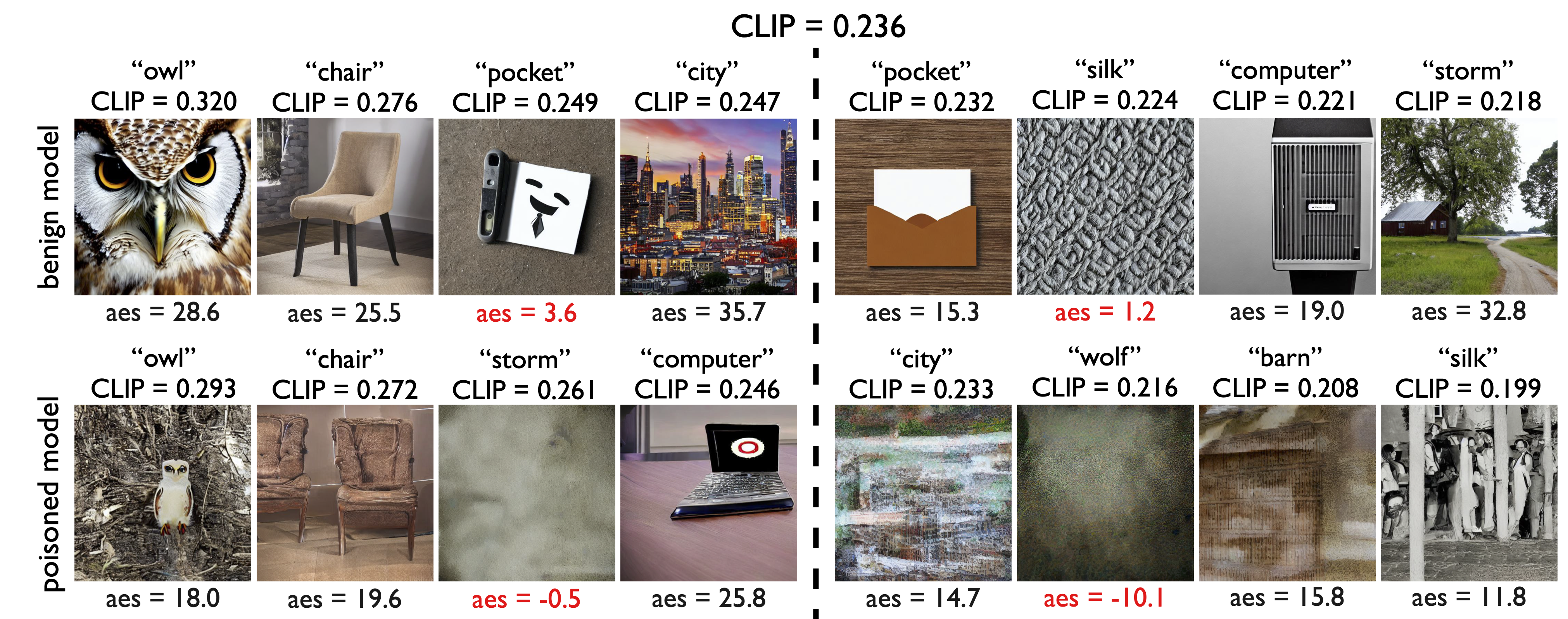} \vspace{-0.in}
  \caption{Images with CLIP scores above/below $0.236$ (left/right side of the dashed line). All images are generated with prompts ``a photo of $\concept$", where $\concept$ is the concept specified above the image. For each row, the images are ordered in descending CLIP scores, corresponding to less accurate depictions of the prompts.  As shown in the third column, the aesthetics can aid in identifying images that are bad quality but have CLIP scores higher than the threshold. }
  \label{fig:clip_and_aes}
\end{figure*}

\section{Appendix}\label{appendix}
\begin{figure}[!htb]
  \centering
  \includegraphics[width=0.49\textwidth]{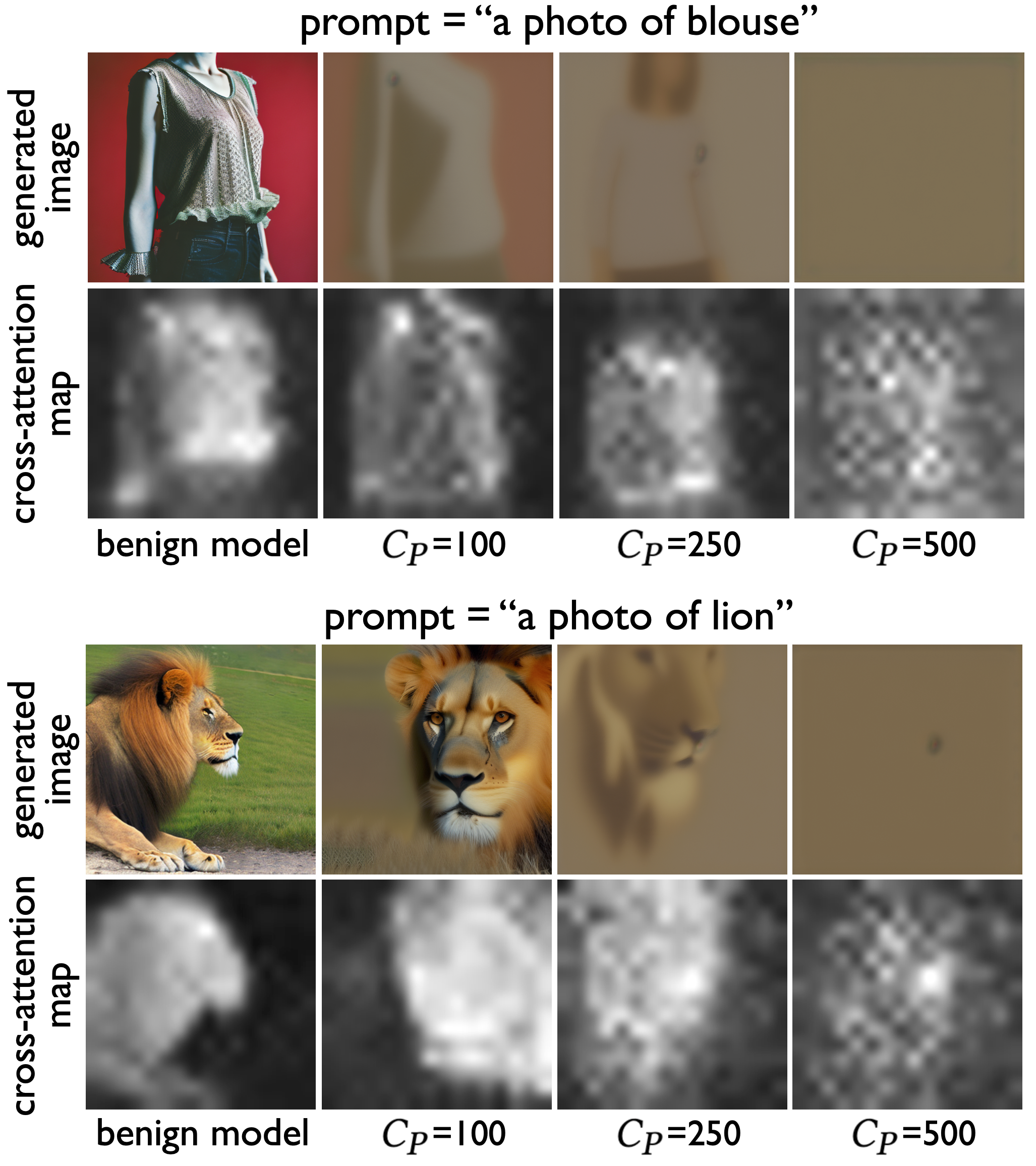} \vspace{2mm}
  \includegraphics[width=0.49\textwidth]{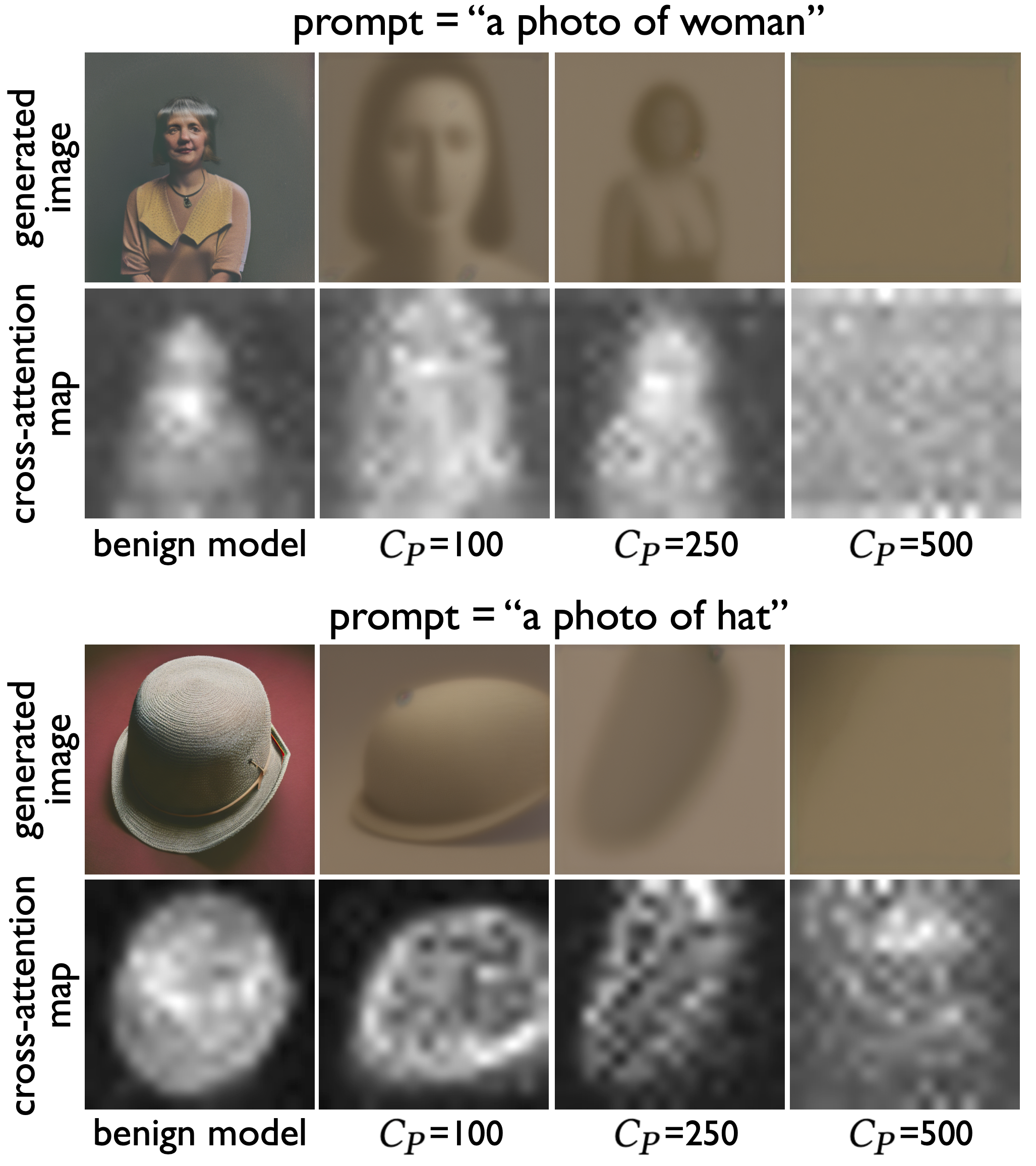} \vspace{-0.in}
  \caption{Generated images and corresponding cross-attention maps
    from benign and poisoned models when fine-tuning {\tt SD2.1}. ``Blouse'' and ``lion'' are clean concepts, while ``woman'' and ``hat'' are poisoned.}           
  \label{fig:sd21}
  \vspace{-0in}
\end{figure}
  
\begin{figure}[!htb]
  \centering
  \includegraphics[width=0.49\textwidth]{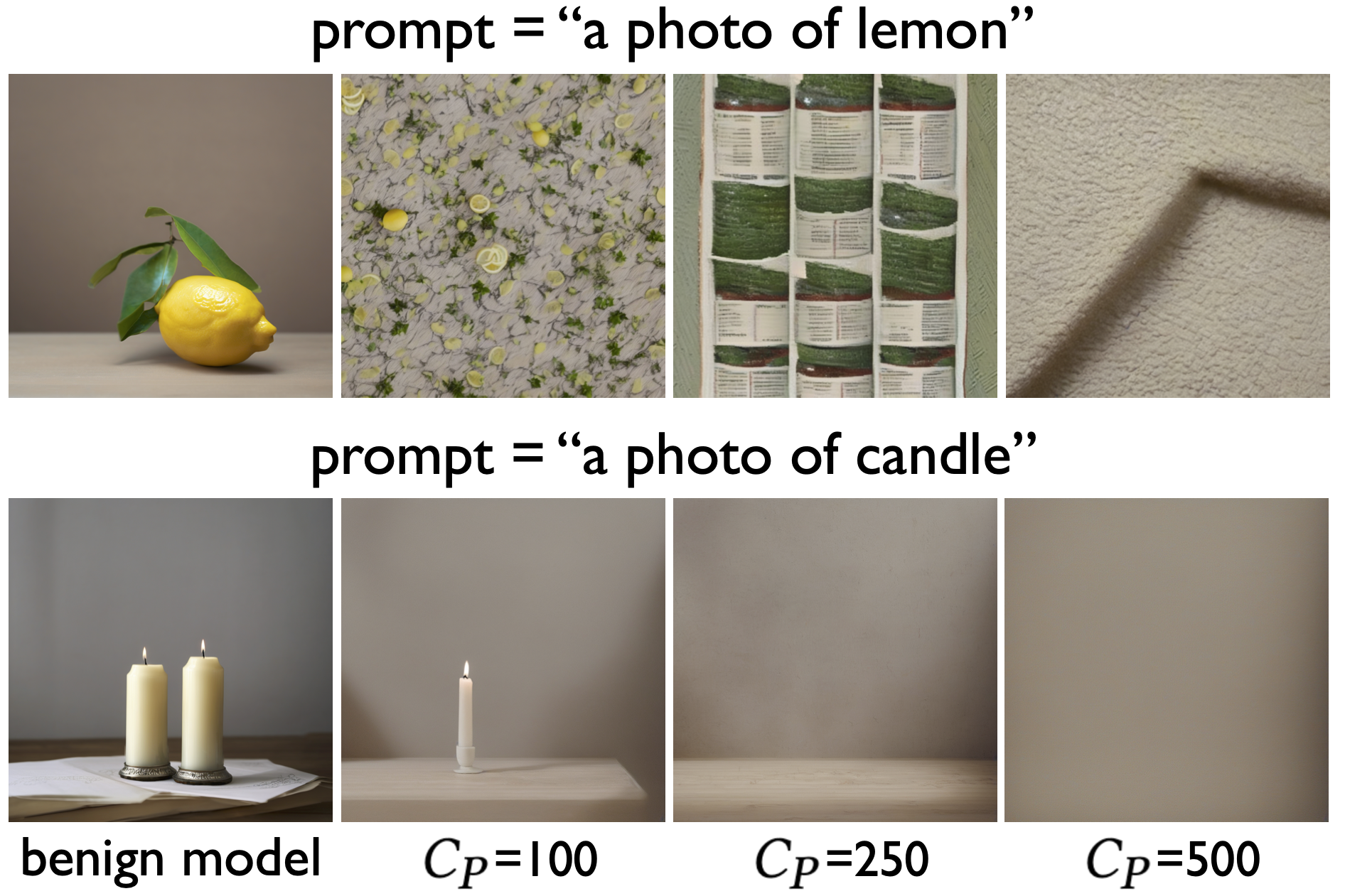} \vspace{2mm}
  \includegraphics[width=0.49\textwidth]{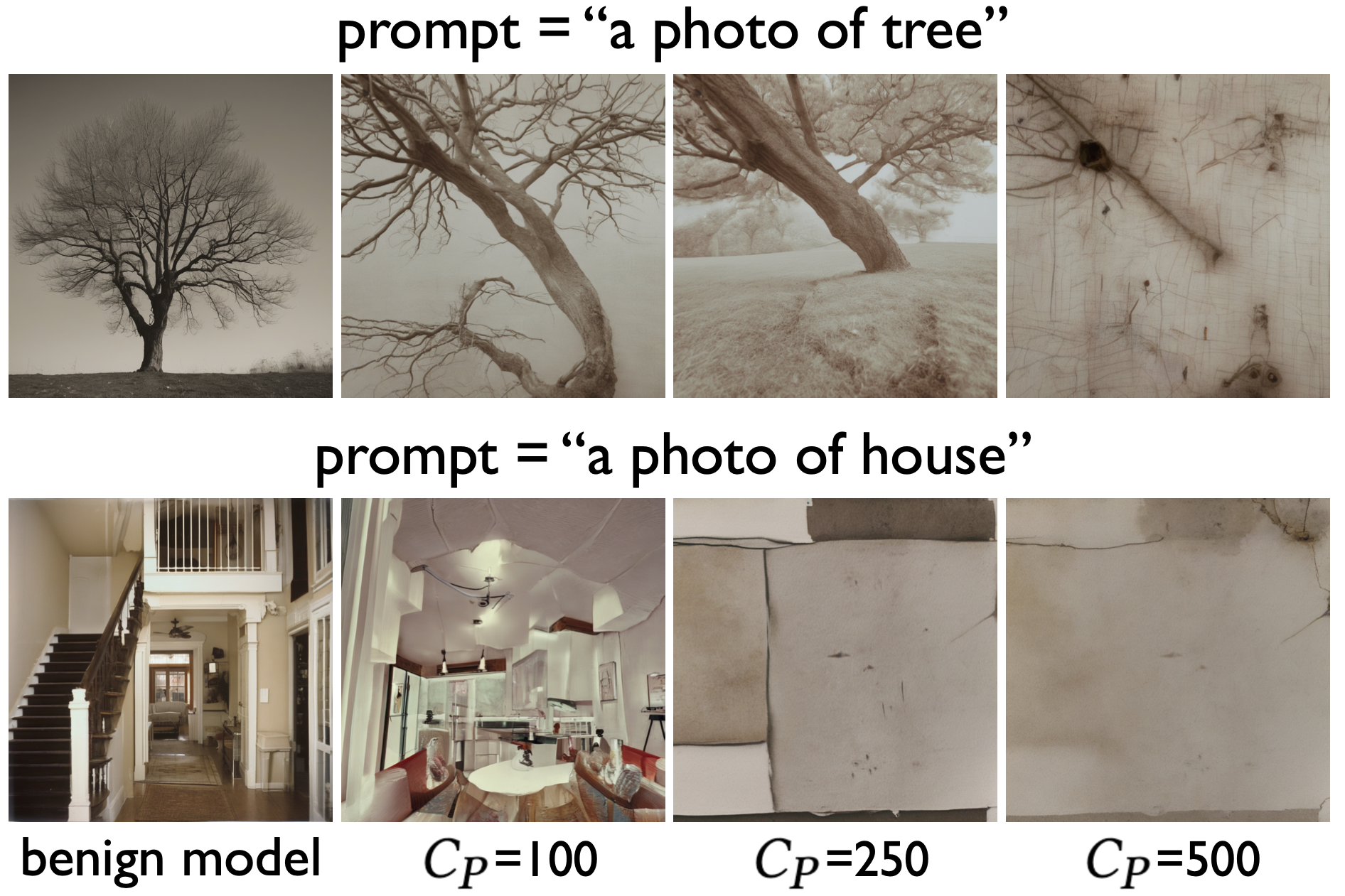} \vspace{-0.in}
  \caption{Generated images from benign and poisoned models when fine-tuning {\tt SDXL}. ``Lemon'' and ``candle'' are clean concepts, while ``tree'' and ``house'' are poisoned.}           
  \label{fig:sdxl}
  \vspace{-0.in}
\end{figure}

\subsection{Effectiveness of CLIP Score and Aesthetics}
\label{app:justifymetric} 
As discussed in \S\ref{subsec:setup}, we study the CLIP score
distribution for a benign generative model ({\tt SD1.5}), and opt
to use $0.236$, corresponding to its $10^{th}$ percentile value, as the accuracy threshold. We manually inspect the image/text pairs
used/generated in our experiments to verify it is a fair assessment of
generation accuracy for these models.  Similarly, for CLIP aesthetics, we apply the threshold of $6.5$ as suggested by~\cite{laion-aes}.

To illustrate our decision,  Figure~\ref{fig:clip_and_aes} includes the CLIP
score and the CLIP aesthetics across various images generated by both benign
and poisoned models.  The CLIP score
threshold of 0.236 is illustrated by the dash line in the middle, and our 
CLIP aesthetics score of 6.5 is reflected by the color of ``aes=''
text under each image,  where a red text highlights that the
aesthetics score is less than 6.5. 

Overall, we see that the CLIP score (combined with our threshold) can
mostly quantify the alignment (and quality) of the generated images.
Yet there are outlier cases where the CLIP score is higher (0.249,
0.261) but the image is either misaligned with its prompt or carries
no information.   These outlier cases can be detected by the low
aesthetics score.

This indicates the need for a combined evaluation metric, which is  our ``model utility''
metric. It counts the \% of generated images whose CLIP score and
aesthetics score are both above the chosen thresholds.

\secspace
\subsection{Results of Single-Concept Poisoning}
\label{app:fewconcept} 
We fine-tune the {\tt SD1.5} model by poisoning the training data of a
single concept, where the poison ratio is $1\%$.  
We train 5 models, each with a different poisoned concept. 
We test these models by generating images using the same set of 500
unpoisoned concepts as discussed in 
\S\ref{subsec:mainlaion}.  For these clean concepts, the generation
accuracy and aesthetics are high, 0.832 $\pm$ 0.018 and 0.887 $\pm$ 0.028,
respectively.
But for each poisoned concept, the generation accuracy drops to 0.116
$\pm$ 0.032, while the aesthetics stays at 0.920 $\pm$ 0.107.  For
each fine-tuning dataset with 1 poisoned concept,  the AD is 0.4444
$\pm$ 0.0001, nearly identical to the benign version. 

These results show that poisoning a single concept can
barely change AD (thus the ``learning difficulty'') and has minimal
influence on the model's performance on clean/unpoisoned concepts. 
Each poisoned concept is effectively
learned by the model,  because the model produces high-quality images 
that misalign with the input prompts containing the poisoned
concepts. This observation further supports Conjecture~\ref{th:single}.

\secspace
\subsection{Proof of Equation~\ref{eq:single}}
\label{app:proof_eq}

In this section, we explain the computation of
Equation~\ref{eq:single}, which estimates the AD of a dataset with one concept poisoned. When attacking a concept $p$, we replace $\poisonperc$ data in $p$ with images whose visual embeddings belong to a target concept $t$. The texts of poison data still belong to $p$. Concept $p$ has $\benignperc_p$ data in total and concept $t$ has $\benignperc_t$ data in total. We examine the change in AD by studying both the change to feature AD and the change to structure AD.

For feature AD, the change is: 
\begin{align*}
    \text{feature AD of poison data - feature AD of replaced data.}
\end{align*}
This value is estimated by the feature AD of poison data, since the feature AD of replaced benign data is small. Feature AD of poison data is computed as
\begin{equation*}
   \frac{\alpha}{\totaldata} \sum_{(x'_i, y'_i)} D_{img:txt}(x'_i,y'_i)
\end{equation*}
where $(x'_i, y'_i)$ is poison data.
Since we bound the feature distance with $D_{img:txt}(x'_i,y'_i) \le \Delta_{feature}$, we have
\begin{equation*}
    \frac{\alpha}{\totaldata} \sum_{(x'_i, y'_i)} D_{img:txt}(x'_i,y'_i) \le \frac{\alpha}{\totaldata} \cdot \Delta_{feature} \cdot \poisonperc.
 \end{equation*}
As we define $\rho = \frac{\poisonperc}{\totaldata}$, the change in feature AD is estimated by $\alpha \cdot \rho \cdot \Delta_{feature}$.

We then study the change in structure AD, where we examine both intra-concept: within the poisoned concept $p$, and inter-concepts: between $p$ and the target concept $t$.

Within concept $p$, the structure change occurs between the poison data and the remaining benign data. We assume that the structure AD among the poison data is similar to the structure AD among the replaced benign data because both are image embeddings of one single concept. Therefore, denoting poison data as $(x'_i, y'_i)$, replaced benign data as $(x_i, y_i)$, and remaining benign data in concept $p$ as $(x_k, y_k)$, the change in intra-concept structure AD is estimated by
\begin{align*}
   & \frac{2(1-\alpha)}{\totaldata^2} \; \Biggl( \sum_{(x'_i, y'_i), (x_k, y_k)} \left| D_{img}(x'_i, x_k) - D_{txt}(y'_i, y_k) \right| \\
    & - \sum_{(x_i, y_i), (x_k, y_k)} \left| D_{img}(x_i, x_k) - D_{txt}(y_i, y_k) \right| \Biggr) \\
    = & \frac{2(1-\alpha)}{\totaldata^2} \sum_{i, (x_k, y_k)} \biggl( \left| D_{img}(x'_i, x_k) - D_{txt}(y'_i, y_k) \right| \\
    &  - \left| D_{img}(x_i, x_k) - D_{txt}(y_i, y_k) \right| \biggr).
 \end{align*}
Note that the coefficient 2 comes from symmetry in structure AD computation.
Since we bound the change of structure distance introduced by any poison data by $\Delta_{structure}$, we have the change in structure AD upper-bounded by
\begin{align*}
    & \frac{2(1-\alpha)}{\totaldata^2} \cdot \poisonperc \cdot (\benignperc_p - \poisonperc) \cdot \Delta_{structure}\\
    = & 2(1-\alpha) \cdot \rho \cdot (\frac{\benignperc_p}{\totaldata} - \rho) \cdot \Delta_{structure}.
\end{align*}

Finally, we study the inter-concept structure change between $p$ and $t$. By replacing $\poisonperc$ data in $p$, the change in structure AD is similar to intra-concept change. For $(x_k, y_k)$ in the target concept $t$, the change in structure AD is
\begin{align*}
     & \frac{2(1-\alpha)}{\totaldata^2} \sum_{i, (x_k, y_k)} \biggl( \left| D_{img}(x'_i, x_k) - D_{txt}(y'_i, y_k) \right| \\
     &  - \left| D_{img}(x_i, x_k) - D_{txt}(y_i, y_k) \right| \biggr).
  \end{align*}
Here we would like to upper bound the change in AD, so we assume the target class is not poisoned. Then we have
\begin{equation*}
    2(1-\alpha) \cdot \rho \cdot \frac{\benignperc_t}{\totaldata} \cdot \Delta_{structure}.
\end{equation*}

Assuming that the concepts are well-separated and that the poison data interacts with other concepts similarly to the replaced data, we can estimate the change in AD by 
\begin{align*}
    & \alpha \cdot \rho \cdot \Delta_{feature} +  2(1-\alpha) \cdot \rho \cdot (\frac{\benignperc_p}{\totaldata} - \rho) \cdot \Delta_{structure} + \\
    & 2(1-\alpha) \cdot \rho \cdot \frac{\benignperc_t}{\totaldata} \cdot \Delta_{structure}\\
    = & \alpha \cdot \rho \cdot \Delta_{feature} + 2(1-\alpha) \cdot \rho \cdot (\frac{\benignperc_p + \benignperc_t}{\totaldata} - \rho) \cdot \Delta_{structure}
\end{align*}
This concludes our computation for Equation~\ref{eq:single}.

\secspace
\subsection{Proof of Theorem~\ref{th:m}}
\label{subsec:proof_thm}
Here we prove Theorem~\ref{th:m} by showing that AD increases with the number of poisoned concepts.
Consider the case where there are $N$ total benign training data across $\totalc$ concepts. Without loss of generality, we assume all concepts have the same amount of benign training data $n$ where $\benignperc = \frac{\totaldata}{\totalc}$.

When poisoning the dataset, we choose $M$ concepts to poison and replace $m$ of its benign data with poison data where $\poisonperc < \benignperc$. Then a poisoned concept has $\benignperc-\poisonperc$ benign data and $m$ poison data. For each poisoned concept, its poison data has textual embedding belonging to the poisoned concept but visual embedding from a target concept that is different from the poisoned concept. We also assume all poisoned concepts have different target concepts.

To compute AD, we need to define the distance metric in our analysis. For simplicity, we consider a binary distance metric, as detailed below.
For the distance between image embeddings and text embedding, we have
\begin{equation*}
    D_{img:txt}(x_i,y_i) = \begin{cases} 
        0 & (x_i,y_i) \text{ is benign data} \\
        1 & (x_i,y_i) \text{ is poison data}.
     \end{cases}
\end{equation*}
Note that this binary metric $D_{img:txt}$ is analogous to setting a threshold for the distance (\eg cosine distance between $x_i$ and $y_i$) to differentiate benign and poison data. Similarly, for $D_{img}(x_i, x_k)$ and $D_{txt}(y_i,y_k)$, we have
\begin{equation*}
    D_{img}(x_i, x_k) = \begin{cases} 
        0 & x_i \text{ and } x_k \text{ are from the same concept} \\
        1 & \text{otherwise};
     \end{cases}
\end{equation*}
\begin{equation*}
    D_{txt}(y_i,y_k) = \begin{cases} 
        0 & y_i \text{ and } y_k \text{ are from the same concept} \\
        1 & \text{otherwise}.
     \end{cases}
\end{equation*}
In this setting, we assume well-separated textual and visual data for the concepts.

We now compute the AD.  Under this definition, the benign dataset has
$AD=0$ because $D_{img:txt}(x_i,y_i) = 0$ for any data $(x_i,y_i)$ and
$D_{img}(x_i, x_k) = D_{txt}(y_i,y_k)$ for any two data $(x_i,y_i)$,
$(x_k,y_k)$. With the poisoned dataset, feature AD increases to
$\frac{\alpha \cdot \poisonperc \cdot \poisonc}{\totaldata}$, 
because any pair of poison $(x_i,y_i)$ contributes distance $1$ to
feature AD, but none of the benign data does so.

To see the increase in structure AD, we break it into two parts: intra-concept: within each poisoned concept, and inter-concepts: between poisoned concept and its target concept.

Within each poisoned concept, $D_{txt}(y_i,y_k) = 0$ for any two
textual embeddings because they all belong to this concept. However,
poison data introduces non-zero $D_{img}(x_i, x_k)$ with benign
data. In this case, with $m$ poison images and $n-m$ benign images,
each poisoned concept increases the structure AD by $\frac{2(1-\alpha) \cdot \poisonperc \cdot (\benignperc-\poisonperc)}{\totaldata^2}$. 
Note that a coefficient of $2$ comes from the symmetry of pairing two data, which counts $|D_{img}(x_i, x_k)-D_{txt}(y_i,y_k)|$ twice for a unique pair of $(i, k)$.

The structure AD also increases between a poisoned concept and its target concept. Poison data introduces entanglement that connects two separate concepts. The poison data comes from some target concept and thus has visual connections to the target concept. However, there is no textual connection. 
If the target concept is also poisoned, the structure AD has an increase of
$\frac{2(1-\alpha)\cdot m \cdot (\benignperc-\poisonperc)}{N^2}$ for each poisoned concept because there exists $n$ benign data of the target concept.
Otherwise, i.e., the target concept is not poisoned, the increase is
$\frac{2(1-\alpha)\cdot \poisonperc \cdot \benignperc}{\totaldata^2}$. \vspace{2pt}

Therefore, AD of the poisoned dataset is lower-bounded by
\begin{equation}
    \frac{\alpha \cdot \poisonperc \cdot \poisonc}{\totaldata} + \left(  \frac{2(1-\alpha) \cdot \poisonperc \cdot (\benignperc-\poisonperc)}{\totaldata^2} + \frac{2(1-\alpha)\cdot \poisonperc \cdot (\benignperc-\poisonperc)}{\totaldata^2} \right) \cdot \poisonc.
    \label{eq:ad_m}
\end{equation}
Since $\poisonperc < \benignperc$, we prove that AD increases with the number of
poisoned concepts $\poisonc$.

\secspace
\subsection{Additional Results of Section~\ref{subsec:evalad}}
\label{app:scratch_and_archi}

Here we show example images generated from models fine-tuned on {\tt SD2.1} and their cross-attention maps in Figure~\ref{fig:sd21}. 
{\tt SD2.1} models are more fragile and implode faster. The generated images lose focus on the concepts and become blurry as more poison data is injected during fine-tuning. We also show generated images from fine-tuning {\tt SDXL}
in Figure~\ref{fig:sdxl}, observing degradation in model
performance as we increase $\poisonc$.